\documentclass[10pt]{iopart}
\usepackage[utf8]{inputenc}
\usepackage{amsfonts}
\expandafter\let\csname equation*\endcsname\relax
\expandafter\let\csname endequation*\endcsname\relax
\usepackage{amsmath}
\usepackage{amssymb}
\usepackage{xfrac}
\usepackage{graphicx}
\usepackage{soul}
\usepackage[dvipsnames]{xcolor}
\usepackage[toc]{appendix}
\usepackage[symbol]{footmisc}
\renewcommand{\thefootnote}{\fnsymbol{footnote}}
\newcommand{\gene}{\textsc{GENE}}
\newcommand{\geneqlnn}{\textsc{GENE}-QLNN}

\newcommand{\rlti}{$R/L_{Ti}$}
\newcommand{\rlte}{$R/L_{Te}$}
\newcommand{\rlne}{$R/L_{ne}$}

\usepackage[backend=biber,maxcitenames=10,sorting=none,style=phys,natbib=true]{biblatex}
\bibliography{bibfile}
\begin{document}


\title{Fast transport simulations with higher-fidelity surrogate models for ITER}

\author[J. Citrin$^{1,2,\dagger}$, P. Trochim$^{3,*}$, T. Goerler$^4$, D. Pfau$^3$, K.L. van de Plassche$^1$ and F. Jenko$^4$]{J. Citrin$^{1,2,\dagger}$, P. Trochim$^{3,*}$, T. Goerler$^4$, D. Pfau$^3$, K.L. van de Plassche$^1$ and F. Jenko$^4$}
\address{$^1$ DIFFER - Dutch Institute for Fundamental Energy Research, 5612 AJ Eindhoven, the Netherlands}
\address{$^2$  Science and Technology of Nuclear Fusion Group, Eindhoven University of Technology, 5612 AZ Eindhoven, Netherlands}
\address{$^3$ DeepMind, London N1C 4AG, UK}
\address{$^4$ Max-Planck Institut für Plasmaphysik, D-85748 Garching, Germany}
\vspace{2mm}
\address{$^\dagger$ Corresponding author: J.Citrin@differ.nl}
\let\thefootnote\relax\footnotetext{~Current affiliation is Meta Platforms, Inc., USA. All work on this project carried out while at DeepMind}
\begin{abstract} %
A fast and accurate turbulence transport model based on quasilinear gyrokinetics is developed. The model consists of a set of neural networks trained on a bespoke quasilinear {\gene} dataset, with a saturation rule calibrated to dedicated nonlinear simulations. The resultant neural network is approximately eight orders of magnitude faster than the original {\gene} quasilinear calculations. ITER predictions with the new model project a fusion gain in line with ITER targets. While the dataset is currently limited to the ITER baseline regime, this approach illustrates a pathway to develop reduced-order turbulence models both faster and more accurate than the current state-of-the-art.  
\end{abstract}

\maketitle
\ioptwocol

\section{Introduction}
Accurate predictive modelling of tokamak turbulence is a key component of multi-physics (integrated) tokamak simulators~\cite{poli:2018}, which are applied for physics interpretation of present-day experiments, and the prediction, optimization and design of future scenarios and devices. The gyrokinetic framework has proven to be successful in quantitatively describing tokamak core turbulence~\cite{garbet:2010,white:2019}. Due to the computational burden in directly applying nonlinear gyrokinetic predictions within integrated modelling, theory-based reduced-order models have been developed, invoking the quasilinear assumption, justified across broad parameter regimes~\cite{staebler:2022}. The essence of the quasilinear assumption is that plasma fluctuations in the nonlinear turbulent state maintain the characteristics of the underlying linear modes, at least at transport-driving scale lengths. Thus, a wavenumber spectrum of the turbulent fluxes can be calculated in part by solving the linear dispersion relation of the underlying system. The saturated amplitude and spectral form of the fluctuations arise from nonlinear physics, and these are captured using a saturation rule in the quasilinear model, with coefficients calibrated to databases of nonlinear gyrokinetic simulations. However, a quasilinear turbulence model with the linear physics calculated by high-fidelity gyrokinetic codes is still too slow for routine application in multiphyics (integrated) tokamak simulation suites. Therefore, reduced-order quasilinear turbulence models have been developed, such as TGLF~\cite{staebler:2007}, QuaLiKiz~\cite{bourdelle:2015,citrin:2017,stephens:2021}, and the Multi-Mode-Model (MMM)~\cite{rafiq:2014}, which approximate gyrokinetic linear solutions and are approximately six orders of magnitude faster to compute than nonlinear gyrokinetics, and enable time-dependent integrated modelling over multiple energy confinement times on the timescale of hours or days (depending on case) using moderate compute resources (e.g. 16 cores at 3.0GHz). TGLF is a gyrofluid model, comprising of velocity space moments of the linear gyrokinetic equations, with a closure designed to approximate the true kinetic response including wave-particle resonances. QuaLiKiz solves the linear gyrokinetic equation directly, but with simplifications for increased tractability, such as pre-computing an approximated eigenfunction from a high-mode-frequency expansion of the gyrokinetic equation. The MMM model is comprised of the superposition of disparate models, including the Weiland fluid model, a drift-resistive-inertial-ballooning mode model, and a critical-gradient-model for electron temperature gradient modes. However, while useful for many applications, the compute timescale of these reduced models still hampers many-query applications related to optimization, design and control. 

The development of surrogates of quasilinear turbulent transport models using machine learning (ML) techniques is a powerful method for model acceleration to facilitate such use-cases. Supervised learning models are trained on pre-generated databases of transport model runs in relevant parameter space. These models, such as neural networks (NNs), have sufficient generality to capture the nonlinear input-output mapping of the original model. Inference times are orders of magnitude faster, on the sub-ms scale, opening up applications such as large-scale validation, uncertainty quantification (UQ), scenario optimization (including inter-shot), controller design, and machine design. 

Following an initial proof-of-principle~\cite{citrin:2015,felici:2018}, NN surrogates of the gyrokinetic quasilinear transport model QuaLiKiz were developed for general~\cite{vandeplassche:2020}, and JET-specific~\cite{ho:2021} input parameters. Recent applications within integrated modelling include optimization studies for the ITER hybrid scenario~\cite{vanmulders:2021}, JET current ramp-up~\cite{ho:2023} and DTT design~\cite{casiraghi:2021}. NN surrogates of TGLF~\cite{meneghini:2017,meneghini:2020} and MMM~\cite{morosohk:2021} have also been developed, with demonstrated applications in core-edge coupled ITER simulations~\cite{meneghini:2020} and control-oriented DIII-D modelling~\cite{morosohk:2020}. Beyond turbulent transport, NN models aiming to accelerate integrated modelling were developed for theory-based~\cite{meneghini:2020} and data-driven~\cite{gillgren:2022} pedestal modelling, and neutral beam heating~\cite{boyer:2019}. Additional NN surrogates were developed for MagnetoHydroDynamic (MHD) stability within the context of disruption prediction~\cite{piccione:2020}, and 3D MHD equilibrium~\cite{merlo:2021}.

While accurate in many standard regimes, the existing reduced-order turbulence models are not fully validated against higher-fidelity models across tokamak plasma parameter space. Known challenges include: resistive-drift modes in the L-mode near-edge~\cite{bonanomi:2021}, which may be important for accurate current ramp-up modelling and first-principle based L--H transition modelling; Kinetic Ballooning Modes (KBM) in the inner core~\cite{kumar:2021}, important for full-profile predictions in MHD-free high-performance hybrid scenarios; fast-ion-enhanced electromagnetic stabilization of Ion Temperature Gradient (ITG) turbulence~\cite{citrin:2013,casson:2020,reisner:2020}, important for high-performance reactor regimes; inter-ELM (Edge Localized Mode) turbulence in the pedestal~\cite{pueschel:2019}, important for prediction of ELM-free regimes as well as accurate pedestal-top boundary conditions for density and temperatures; electromagnetic microtearing turbulence~\cite{jian:2021}, important in high $\beta_p$ regimes and pedestals; and spherical tokamak turbulence regimes~\cite{patel:2019}. The model deficiencies may be due to fundamental assumptions inherent to the present implementations of the models, as well as challenges in setting universally appropriate model parameters. Furthermore, even in standard regimes, the linear modes calculated by the reduced-order models are not perfect reproductions of higher-fidelity gyrokinetic linear solvers. 

Continuous refinement of the reduced-order-models may further close the gap to high-fidelity models. However, another approach is to directly train a surrogate on a higher-fidelity model, whether nonlinear, or quasilinear using a high-fidelity (with respect to linear physics) gyrokinetic code, taking advantage of the fact that the compute time required is primarily for training set generation. Provided enough High Performance Computing (HPC) capacity, training sets with sufficient parameter space coverage can be generated using high-fidelity models not routinely used within integerated modelling, due to their computational expense. A surrogate trained on the higher-fidelity model would result in a physics model both faster and more accurate than current state-of-the-art quasilinear transport models such as TGLF and QuaLiKiz, dependent on sufficient quality training sets and appropriate surrogate model training choices. Since, as yet, nonlinear simulations are too expensive to be the sole source of training set data, we propose a multi-fidelity approach. A database of quasilinear transport fluxes is built, based on linear gyrokinetic simulations and a nonlinear saturation rule, calibrated by limited nonlinear simulations strategically spanning the parameter space. The advantage over present-day tokamak turbulent transport model NN surrogates is that the underlying linear modes in the database are calculated with minimal assumptions and are thus the highest achievable fidelity, including in the challenging regimes listed previously. These calculations, which are the most computationally expensive component of a quasilinear model, thus remain relevant for perpetuity, and are applicable wherever the quasilinear assumption is valid. The quasilinear model is then modular in the sense that various saturation rules can be applied to the linear mode database with minimal computational expense, and tested in integrated modelling. The saturation rules are expected to improve and evolve, through both new approaches~\cite{staebler:2021,dudding:2022}, and the expanding availability of nonlinear simulations in relevant parameter space, as we do here.

In this paper, we demonstrate a proof-of-principle of a heat and particle transport {\gene}~\cite{jenko:2000b} quasilinear (QL) NN transport model. We initially focus on parameters corresponding to the ITER baseline scenario taken from existing QuaLiKiz and TGLF extrapolations of ITER performance~\cite{mantica:2020}. Linear {\gene} runs are carried out based on variations of input parameters taken from the final state of the integrated modelling QuaLiKiz simulation in Ref.~\cite{mantica:2020}. Dedicated nonlinear runs at mid-radius were then carried out, to validate a custom saturation rule developed for this regime. The resultant QL transport flux database was then fit with NN models. Finally, the developed {\geneqlnn} surrogate was coupled to the JINTRAC~\cite{cenacchi:1988,romanelli:2014} tokamak simulator, and the ITER baseline case ($B_T=5.3T, I_p=15MA$) was rerun. Therefore, this work also serves as a higher fidelity extrapolation of ITER performance than presently existing, and a validation test of reduced-order models in this specific regime. 

The approach at hand is similar to the one of the DeKanis project~\cite{narita:2021}, with a number of differences. Our saturation rule incorporates multi-spectral information, and is tuned to theory-based nonlinear gyrokinetic simulations instead of being semi-empirical. Furthermore, our initial application is for ITER extrapolation, which is novel for this class of high-fidelity surrogate turbulence model. Going forward, both projects would benefit from shared linear gyrokinetic datasets.  

We note that ML techniques can also be used to accelerate high-fidelity nonlinear gyrokinetic modelling, as illustrated by a recent application of Gaussian Process Regression for iteratively generating local (in parameter space) surrogate models of a nonlinear gyrokinetic model to accelerate convergence within a flux-balanced stationary-state solver~\cite{rodriguez-fernandez:2022c}. This technique is extremely valuable in accelerating ultra-high-fidelity validation of scenario predictions by a significant factor. However, since nonlinear simulations are still applied within the workflow, the computational time is still prohibitively expensive for time dependent, scenario optimization, and control-oriented modelling, and thus serves a different use-case. 

The rest of this paper is organized as follows. Section~\ref{sec:trainingset} covers the training set generation, which includes {\gene} linear runs, QL model generation with nonlinear runs and saturation rule calibration, and data filtering. Section~\ref{sec:NNtraining} summarizes the NN training pipeline and surrogate model generation. Implementation within integrated modelling is discussed in section~\ref{sec:integratedmodelling}. Conclusions and outlook are provided in section~\ref{sec:conclusions}. 

\section{Training set generation}
\label{sec:trainingset}
All linear instability calculations were carried out with the {\gene} gyrokinetic model. {\gene} is a Eulerian gyrokinetic code, evolving the perturbed particle distribution functions self-consistently with the Maxwell field equations. {\gene} works in field aligned coordinates, where $x$ is the radial coordinate, $z$ the parallel coordinate along the field line, and $y$ the binormal coordinate. All shown simulations are spectral in both the $x$ and $y$ directions. 

\begin{figure}[hbt]
	\centering
	\includegraphics[width=1.0\linewidth]{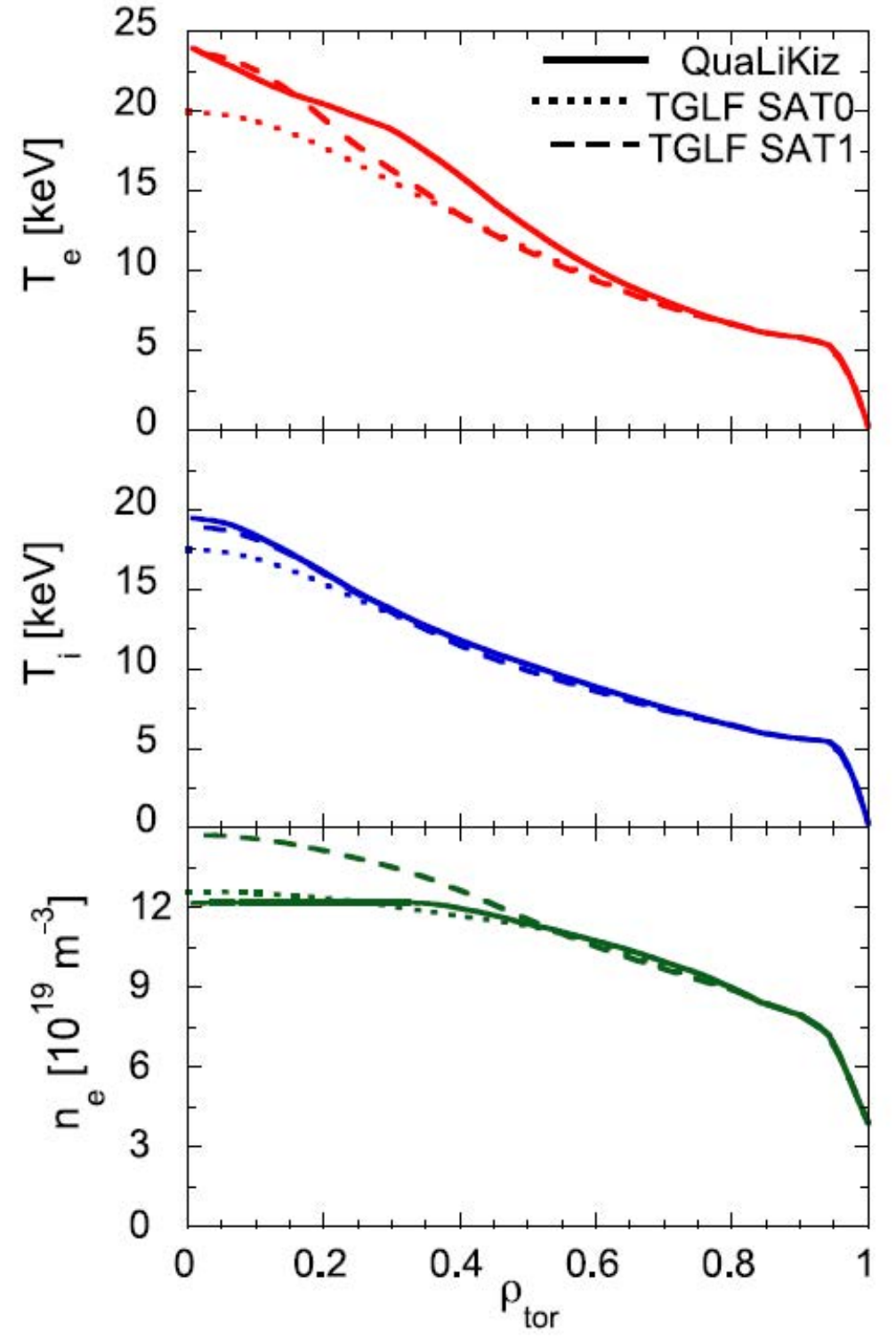}
	\caption{ITER baseline modelling with TGYRO-TGLF-SAT0 (dotted curves), TGYRO-TGLF-SAT1 (dashed curves) and JINTRAC-QuaLiKiz (solid curves). With a pedestal boundary condition at $\rho_{tor}=0.93$, predictions are for $T_e$ (upper panel), $T_i$ (middle panel) and $n_e$ (lower panel). The pedestal boundary condition is consistent with EPED~\cite{snyder:2009} pedestal stability predictions. Figure reproduced with permission from Plasma Phys. Control. Fusion 62, 014021 (2019)~\cite{mantica:2020}. Copyright 2019 Institute of Physics.}
	\label{fig:manticaITER}
\end{figure}

The input parameters for the {\gene} linear dataset were based on the final stationary state of the JINTRAC-QuaLiKiz ITER baseline simulation from Ref.~\cite{mantica:2020}. These simulation results, together with the analogous JINTRAC-TGLF simulations, are reproduced for convenience in figure~\ref{fig:manticaITER}. The average fusion gain predicted by the models is $Q\sim9.5$, in line with ITER goals. 

Eight radial locations were chosen for sampling linear gyrokinetic inputs, ranging from normalized toroidal flux coordinate $\rho_N=0.2$ to $\rho_N=0.9$, in steps of $0.1$. The normalized toroidal flux coordinate is defined as $\rho_N\equiv\sqrt{\frac{\Phi(r)}{\pi B_0}}/\sqrt{\frac{\Phi(a)}{\pi B_0}}$, where $\Phi(r)$ is the toroidal flux within radius $r$, $a$ is the tokamak minor radius, and $B_0$ the toroidal magnetic field at the geometric magnetic axis. This choice was justified as follows: the outer radius is in the vicinity of the pedestal-top boundary condition applied in H-mode core integrated modelling; the inner radius is in the vicinity of the region where within, no instabilities are typically predicted in integrated modelling applications due to a combination of non-negligible ion neoclassical transport, and a reduction of TEM drive due to the proportionality of trapped electron fraction with normalized radius $r/a$, which is typically within a few percent of $\rho_N$. On the stability of the inner radial region, exceptions have been reported in electromagnetic (EM) simulations~\cite{kumar:2021}. However, as will be elucidated further in section~\ref{sec:linear}, EM simulations are out of the scope of this study. External (e.g. from NBI torque) perpendicular flow shear ($E \times B$ rotation shear) effects were also not included in this study. Its direct impact on the linear modes is difficult to assess for large-scale automated run generation due to the non-stationarity of the solutions that rotation induces, resulting in an involved analysis procedure that is technically challenging to automate~\cite{dagnelie:2019}. While a model of the direct impact of $E \times B$ shear on the fluxes is feasible to implement for the NN surrogate, as done in Ref.~\cite{vandeplassche:2020}, it was not implemented in this work, motivated by the moderate impact of $E \times B$ shear expected for the ITER baseline scenario according to TGLF-SAT1 simulations~\cite{chrystal:2020}. The lack of both stabilizing EM and $E \times B$ effects means that the results obtained with our NN surrogate in section~\ref{sec:integratedmodelling} are conservative. The impact of fast ions on the turbulence is also not included due to computational expense, even though their impact may be stabilizing~\cite{citrin:2023}.

The original ITER simulation contained 6 ion species. For computational efficiency, we make a bundled ion assumption. The main ions (deuterium and tritium) are combined into a single effective ion. All impurities are bundled into a single effective impurity, with parameter choices that mimic the impact of all impurities. For further details and justification see Appendix A.

In table~\ref{tab:physicsparams}, the main dimensionless parameters relevant for gyrokinetic simulation input, at each sampled radial location, are listed. The list is partial, and for brevity does not include collisionality and flux surface shaping parameters, although these parameters are included in the calculations themselves. Collisionality is calculated using reference temperatures and densities observable in figure~\ref{fig:manticaITER}, and modelled using a linearised Landau-Boltzmann operator. Flux surface shaping is calculated with Miller parameterization~\cite{miller:1998}, and arises from Grad-Shafranov 2D MHD equilibrium calculations carried out in the integrated modelling simulation, using the standard ITER boundary shape with elongation $\kappa_{95}\approx1.7$ and triangularity $\delta_{95}\approx0.4$.  

\begin{table*}
    \caption{List of main dimensionless input parameters for the {\gene} linear runs, based on a JINTRAC-QuaLiKiz ITER simulation, for all sampled normalized toroidal flux coordinates $\rho_N$. {\rlti}$\equiv-R\frac{\nabla T_i}{T_i}$, where $R$ is the tokamak major radius, and $T_i$ the ion temperature. {\rlte}, {\rlne}, and $R/L_{nimp}$ are the analogous normalized logarithmic gradients of the electron temperature, electron density, and bundled impurity density, respectively. $q$ is the magnetic pitch angle (safety factor), $\hat{s}$ is the magnetic shear related to the derivative of the pitch angle. $\alpha_\mathrm{MHD}\equiv -Rq^2\beta'$, where $\beta'$ is the radial derivative of the plasma beta (ratio of kinetic to magnetic pressure), and is related to the Shafranov shift of the flux surfaces. $Z_\mathrm{eff}$ is the effective ion charge $\Sigma_i \frac{n_i}{n_e}Z_i^2$, and $T_i/T_e$ is the ion to electron temperature ratio.}
    \centering
      \begin{tabular}{c|c|c|c|c|c|c|c|c|c}
	    $\rho_N$ & {\rlti} & {\rlte} & {\rlne} & $\hat{s}$ & $q$ & $\alpha_\mathrm{MHD}$ & $R/L_{nimp}$ & $Z_\mathrm{eff}$  & $T_i/T_e$ \\
        \hline
	    0.2 & 4.54 & 2.22 & 0.12 & 0.16 & 0.95 & 0.28 & 0.9 & 1.87 & 0.82  \\
	    
	    0.3 & 4.68 & 3.61 & 0.30 & 0.12 & 1.01 & 0.33 & 1.05 & 1.84 & 0.76 \\
	    
	    0.4 & 4.40 & 5.70 & 1.06 & 0.39 & 1.09 & 0.45 & 2.40 & 1.76 & 0.76 \\
	    
	    0.5 & 4.63 & 7.17 & 1.97 & 0.85 & 1.23 & 0.56 & 5.59 & 1.71 & 0.82\\
	    
	    0.6 & 5.53 & 7.57 & 2.13 & 1.26 & 1.48 & 0.64 & 6.30 & 1.67 & 0.89  \\
	    
	    0.7 & 5.8 & 7.48 & 2.52 & 1.63 & 1.80 & 0.77 & 3.47 & 1.62 & 0.94 \\
	    
	    0.8 & 6.4 & 7.49 & 4.94 & 2.24 & 2.21 & 1.11 & 13.65 & 1.62 & 0.99 \\
	    
	    0.9 & 9.56 & 8.97 & 7.32 & 3.70 & 2.83 & 1.95 & 11.06 & 1.63 & 1.01 
	\end{tabular}
	\label{tab:physicsparams}
	\normalsize
\end{table*}

\subsection{{\gene} linear runs}
\label{sec:linear}

The gyrokinetic inputs for the NN training set are sampled from variations of the parameters listed in table~\ref{tab:physicsparams}. Since the intent is to apply the resultant transport model within integrated modelling for current, heat and particle transport, deviations of the predicted plasma profiles in the subsequent simulation, away from these nominal parameters, are expected. Thus, to avoid NN extrapolation by ensuring that all predictions are encapsulated within the training set envelope, a training set was generated with, at each radial location, a 5-point scan of {\rlti}, {\rlte} and {\rlne}. For each variable, variations of $\pm15\%$ and $\pm30\%$ were carried out. 3-point scans of $\hat{s}$ and $q$ with $\pm15\%$ were also carried out, leading to a total of 1125 calculations for each wavenumber at each radial location. All other dimensionless parameters were kept fixed in these scans, corresponding to their nominal values at each radial location, and are not inputs into the NN surrogate model. We denote this initial scan as Phase 1. Subsequent integrated modelling simulations with the NN model constructed from Phase 1 found this initial scan to be insufficient for capturing the entirety of the resultant predicted parameter space, leading to spurious results. Namely, low {\rlti}, {\rlte} in a specific radial zone was predicted by the transport equations, outside the Phase 1 data boundaries, leading to NN extrapolation with incorrect trends, triggering a physically incorrect feedback mechanism where even lower {\rlti}, {\rlte} was predicted. Thus, in a Phase 2 of the run generation, additional scans were carried out at nominal $\hat{s}$ and $q$, but with increased sampling at low logarithmic gradients, at factors $0.5$ and $0.1$ of the nominal {\rlti}, {\rlte}, and {\rlne} respectively. Such a multi-step approach is also relevant for the general case of surrogate model generation, where a metric of model trustworthiness during application should be fed back into the training set generation pipeline for subsequent model refinement. The following sections of quasilinear model generation (section~\ref{sec:quasilinear}), NN model generation (section~\ref{sec:NNtraining}), and integrated modelling (section~\ref{sec:integratedmodelling}) include data of both Phase 1 and 2. The impact of only including Phase 1 data is shown in Appendix B.

A wider range of sampling around additional input dimensions is desirable. However, considering the computational expense (1-5 CPUh per linear calculation, defined as compuational wall-time multiplied by the number of CPUs in the calculation) and technical challenges of setting up the required automated run generation and analysis pipelines for larger scale parameter scans, this initial {\gene} quasilinear NN model is limited to the aforementioned input dimensions and the ITER baseline regime. 

Linear simulations for nine wavenumbers at ion-Larmor-radius scalelengths were carried out, at normalized wavenumbers $k_y\rho_s=$[0.1, 0.15, 0.2, 0.25, 0.325, 0.4, 0.5, 0.7, 1.0] (henceforth abbreviated as $k_y$), with the reference Larmor radius $\rho_s\equiv\frac{\sqrt{T_em_D}}{q_eB}$. $T_e$ is the electron temperature, $m_D$ the deuterium mass, $q_e$ the electron charge, and $B$ the reference magnetic field at the tokamak magnetic axis. Electron-Larmor-radius scalelengths were deemed unnecessary to include since Electron Temperature Gradient modes (ETG) have been predicted to be subdominant in the ITER regime with $T_e>T_i$, which increases ETG critical thresholds~\cite{mantica:2020}. However, direct multiscale simulation of the ITER regime is still necessary to fully answer this question. Studies focusing on electron heated regimes have also predicted scenarios where multiscale interactions become important, even potentially improving confinement~\cite{maeyama:2022}.

All simulations carried out were electrostatic (ES), meaning that magnetic field fluctuations are neglected by setting $\beta\sim0$. The magnetic geometry $\alpha_\mathrm{MHD}$ parameter was still set according to the nominal $\beta'$. The ES assumption is conservative, since for the $\beta<2\%$ values typical of the ITER baseline scenario, EM-modes are not expected to be destabilized in the plasma core (with the possible exception of the magnetic axis region~\cite{kumar:2021}), while EM-stabilization of ITG turbulence is expected to contribute to improved confinement, and is further enhanced by energetic ion species induced by fusion reactions and the Neutral Beam Injection (NBI) system, neither of which is included in this study~\cite{citrin:2013,citrin:2014b,garcia:2016,garcia:2018,mazzi:2022}. Finite $\beta$ simulations have a greater computational burden, and more challenging numerical convergence and automated mode convergence properties, less conducive for automated large-scale run generation as carried out here. Therefore the generalization to finite-$\beta$ is left for future work. 

In total, for both Phase 1 and 2, approximately 100,000 linear instability calculations were carried out (81,000 in Phase 1 and 15,696 in Phase 2). All calculations were initial value (IV) simulations, whereby following an initial transient phase, the simulation becomes dominated by the fastest (exponentially) growing mode when an instability is present, or conversely by the slowest (exponentially) decaying mode when no instability is present. Ideally, the {\gene} convergence monitoring routine returns either a finite mode growth rate and frequency for the unstable modes, or a zero growth rate when no instability is present. For the unstable modes, all necessary mode characteristics for quasilinear model generation, such as mode spatial structure and transport fluxes, are obtained from the final state of the IV simulation. All runs were carried out on Google Cloud infrastructure, with pipelines constructed for scan generation, run monitoring, result retrieval, and data reduction to the minimum required for generating a quasilinear model and transport flux database. Configurations where all wavenumbers are stable, contribute to a zero-flux (stable) point in the transport flux database. As described in section~\ref{sec:NNtraining}, these stable points are important for capturing sharp critical-gradient thresholds in the resultant NN model. Since the {\gene} convergence monitoring routine does not always identify stable modes, a post-processing tool was developed to identify stable regions and supplement the transport flux database, as described in section~\ref{sec:filtering}.

Numerical convergence, with respect to phase space grid resolution, was determined through dedicated convergence checks on a randomly sampled $\sim1\%$ of the parameter configurations (including wavenumber) at each radial location. Starting from standard grid resolution settings, the grid in each dimension was refined until growth rate convergence was observed. The final grid resolution required at each radial location is displayed in table~\ref{tab:gridparams}. The trends are as expected: at lower radii the modes tend to be less strongly driven and thus higher grid resolution tends to be necessary for a given relative tolerance, together with the fact that more radial modes are necessary at low magnetic shear, reflecting the tendency of modes at low magnetic shear to be less localized in ballooning space. At the larger radius, $n_z$ (parallel) resolution needs to be higher since there is relatively more variation of magnetic geometry parameters along the field line at that position. Note that all radial wavenumbers considered, are coupled to each other through the twist-and-shift parallel boundary condition~\cite{beer:1995}. At the low-field-side midplane, zero ballooning angle ($k_x=0$) is imposed.

\begin{table}
    \caption{Numerical {\gene} grid resolution for the generated linear dataset. The same resolution is applied for all binormal $k_y$ wavenumbers at a given radial location. $n_{kx}$ is the number of radial modes, $n_z$ the number of parallel (along the B-field) grid points, $n_w$ the number of perpendicular velocity grid points (magnetic moments), and $n_v$ the number of parallel velocity grid points.}
    \centering
      \begin{tabular}{c|c|c|c|c}
	    $\rho_N$ & $n_{kx}$ & $n_z$ & $n_w$ & $n_v$ \\
        \hline
	    0.2 & 25  & 24  & 16  & 48  \\
	    0.3 & 25  & 16  & 16  & 32  \\
	    0.4 & 25  & 16  & 16  & 32  \\
	    0.5 & 17  & 16  & 12  & 32  \\
	    0.6 & 17  & 16  & 12  & 32  \\
	    0.7 & 17  & 16  & 12  & 32  \\
	    0.8 & 17  & 16  & 12  & 32  \\
	    0.9 & 17  & 64  & 12  & 32  \\
	\end{tabular}
	\label{tab:gridparams}
	\normalsize
\end{table}

\begin{figure*}[hbt]
	\centering
	\includegraphics[width=1.0\linewidth]{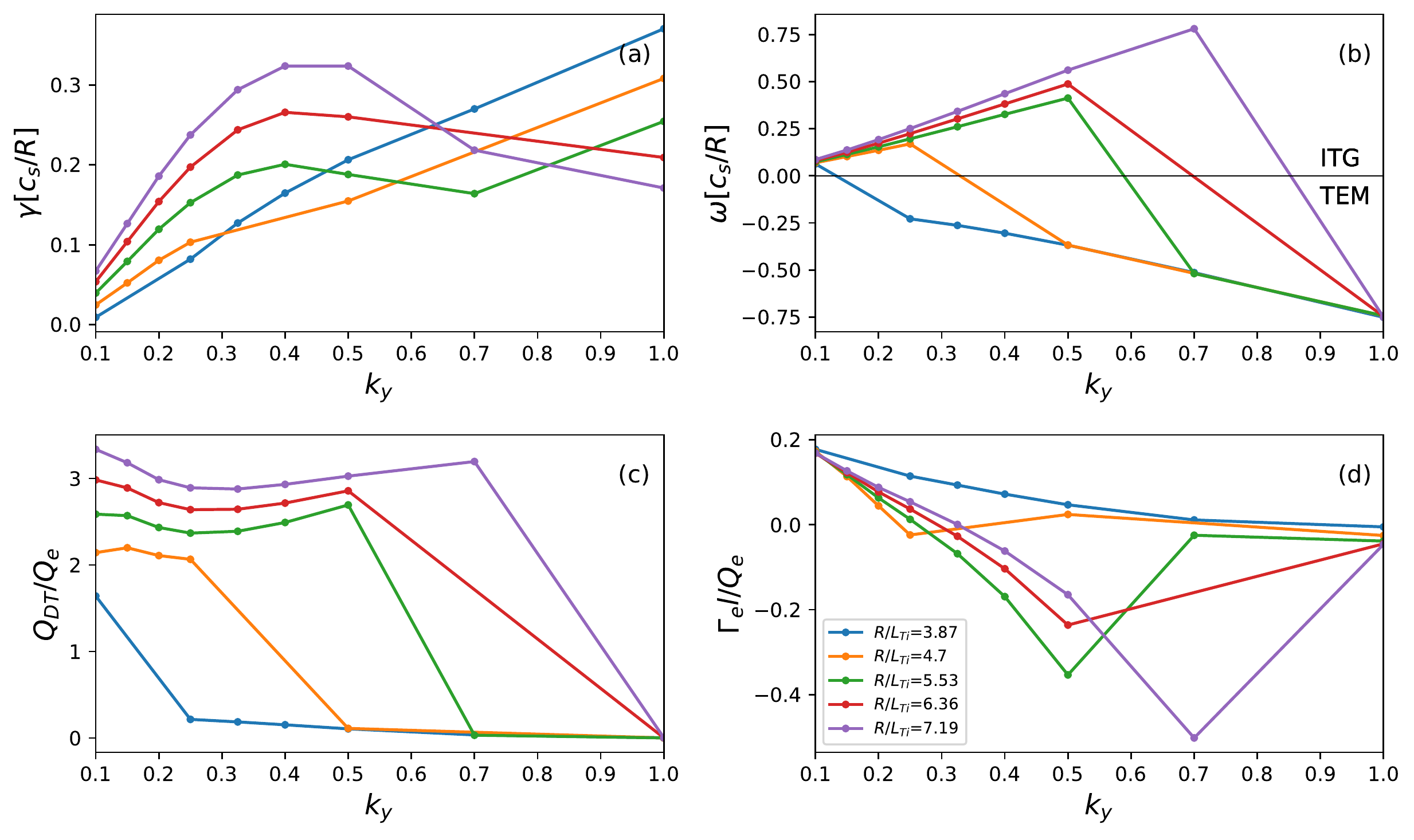}
	\caption{Linear mode spectra and flux ratios, for a typical {\rlti} scan at $\rho_N=0.6$, with {\rlte}=7.57, {\rlne}=2.13, $q$=1.48, and {$\hat{s}$}=1.26. Growth rates (a), mode frequency (b), ion to electron heat flux ratio per mode (c), and electron particle flux to electron heat flux (d) are shown for various values of {\rlti}. Positive frequencies correspond to ITG modes, and negative frequencies to TEMs.}
	\label{fig:linear_rltiscanA}
\end{figure*}

A typical example of mode spectra and transport flux ratios, as needed for quasilinear model generation, is shown in figure~\ref{fig:linear_rltiscanA}. Linear mode eigenvalues (growth rate and frequency) as well as flux ratios per mode are shown for a scan of {\rlti} at $\rho_N=0.6$. In the {\gene} convention, positive frequencies correspond to modes propagating in the ion diamagnetic direction, which are ITG modes in this case. Negative frequencies correspond to modes propagating in the electron diamagnetic direction, TEMs in this case. Typical of electron heated ITER cases, the mode landscape is comprised of a combination of ITG and TEM modes. TEM is more prevalent at lower {\rlti} values in the scan and at higher $k_y$, as observable from the negative mode frequencies and heat flux ratio $Q_{i}/Q_e<1$. Interestingly, there is significant variation in the particle flux in the scan, including transitions from inward to outward transport. This is an indication that particularly in this regime, it is critical for the quasilinear flux model to match the nonlinear flux $k_y$ spectrum, since this has significant impact on particle transport and hence the predicted density peaking. This will be further discussed in section~\ref{sec:quasilinear}.

Only the fastest growing mode is calculated with the IV solver. However, subdominant modes are likely present in this regime, i.e. subdominant TEM when ITG is dominant, and vice versa. Ideally, for a more accurate quasilinear model, these subdominant modes would also be calculated using an eigenvalue (EV) solver in lieu of the IV solver. However, tailoring the EV solver preconditioner across wide parameter space is a complex problem in automated workflows. Therefore this aspect is left for future work. 

\begin{figure}[hbt]
	\centering
	\includegraphics[width=1.0\linewidth]{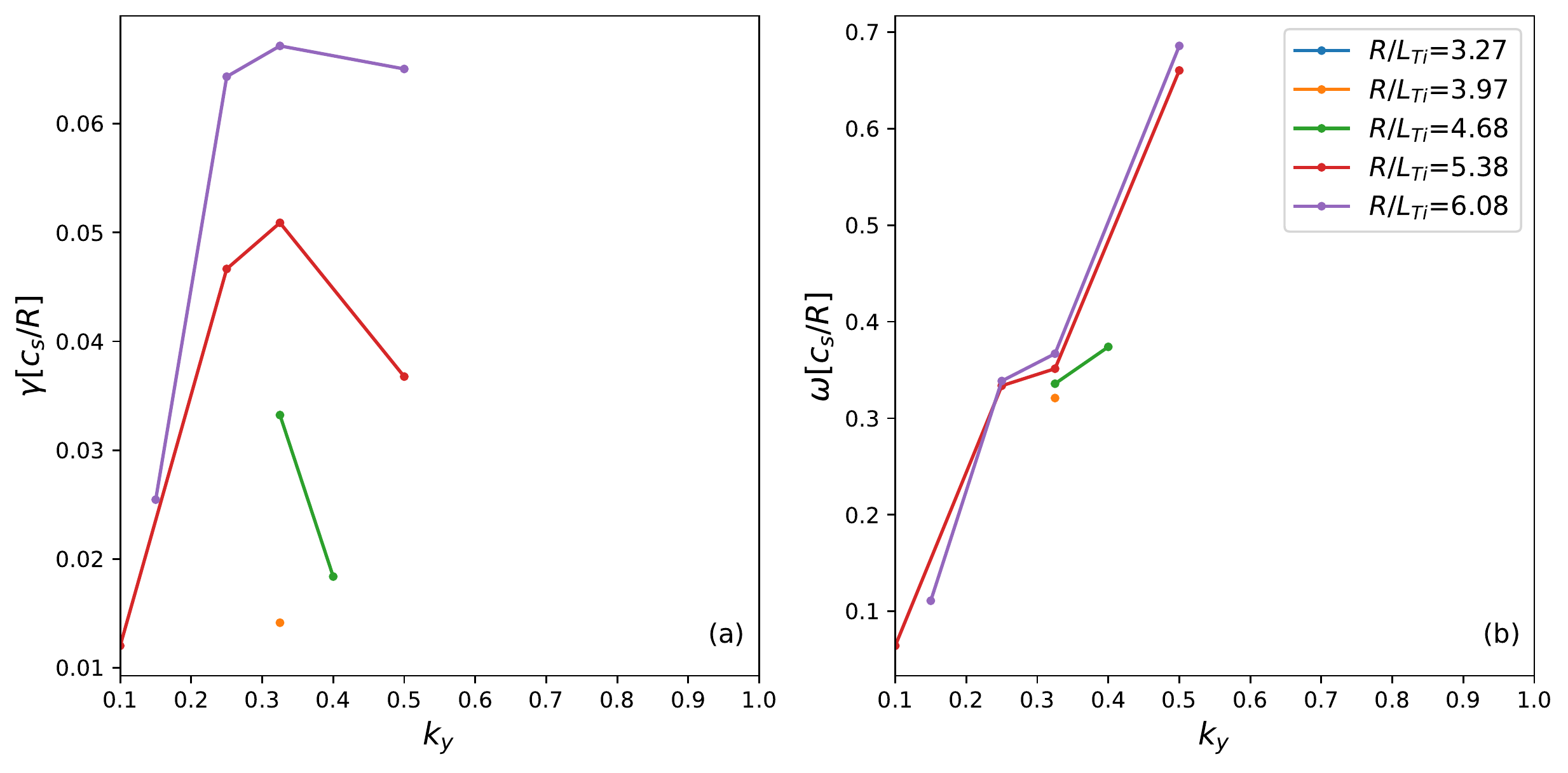}
	\caption{Linear mode spectra for a  {\rlti} scan at $\rho_N=0.3$, with {\rlte}=3.61, {\rlne}=0.3, $q$=1.01, and {$\hat{s}$}=0.12. Growth rates (a) and mode frequency (b) are shown for various values of {\rlti}.}
	\label{fig:linear_rltiscanB}
\end{figure}

Another caveat is the robustness of the {\gene} convergence monitoring routine. Particularly for slowly growing modes just above the instability critical thresholds, as well as for stable modes, there are cases where the {\gene} IV solver does not converge to a stationary solution within the allotted time allocation. While partially mitigated through increasing run time allocations, and reducing the convergence criteria, such cases are still evident in our dataset. While for smaller studies these cases can be resolved through deeper analysis on an individual basis, this is impractical for large-scale studies with automated pipelines. An example is shown in figure~\ref{fig:linear_rltiscanB}, for an {\rlti} scan around a parameter configuration at $\rho_N=0.3$, which is characterised by weakly growing modes. The spectrum at {\rlti}=4.68 likely has a ``missing'' mode at lower $k_y$, due to non-convergence of the solver in this particular case. Since lower $k_y$ is linked with larger transport fluxes (larger fluctuation wavelengths), these missing modes can cause discontinuities in parameter regions near critical thresholds. The ramifications of this data imperfection is further discussed in section~\ref{sec:NNtraining}.

\subsection{Quasilinear model training set generation}
\label{sec:quasilinear}
Using the linear mode dataset, quasilinear fluxes are constructed with the aid of a saturation rule and {\gene} nonlinear simulations carried out in the ITER baseline regime. Following the quasilinear dataset generation, outliers were filtered (due to aforementioned discontinuities) and the dataset was supplemented with zero flux datapoints (stable modes), which also form part of the NN training set. 

\subsubsection{Dedicated nonlinear simulations}
\label{sec:NL}
To calibrate the saturation rule in the specific ITER baseline turbulence regime, dedicated {\gene} nonlinear simulations were carried out on the Marconi HPC cluster. Two separate {\rlti} scans were made, for the parameters corresponding to $\rho_N=0.5$. The first {\rlti} scan was with the nominal parameters from table~\ref{tab:physicsparams}. The second scan was the same apart from {\rlte} and {\rlne} reduced by 30\%, to enter a more ITG dominant regime. Simulation box size and grid convergence was verified through dedicated checks with increased box size and resolution. The applied perpendicular box sizes were $[L_x,L_y]\approx[135,125]$ in units of \textsc{GENE} reference Larmor radius (which is close to the ion Larmor radius), with $[n_{kx}, n_{ky}]=[192,32]$ perpendicular wavenumbers. $n_z=18$ for the parallel grid resolution, and $[n_w,n_v]=[32,12]$ velocity grid points were used. For context, the reference Larmor radius at $\rho_N=0.5$ is $\approx3~mm$ in expected ITER conditions.

\begin{figure}[hbt]
	\centering
	\includegraphics[width=1.0\linewidth]{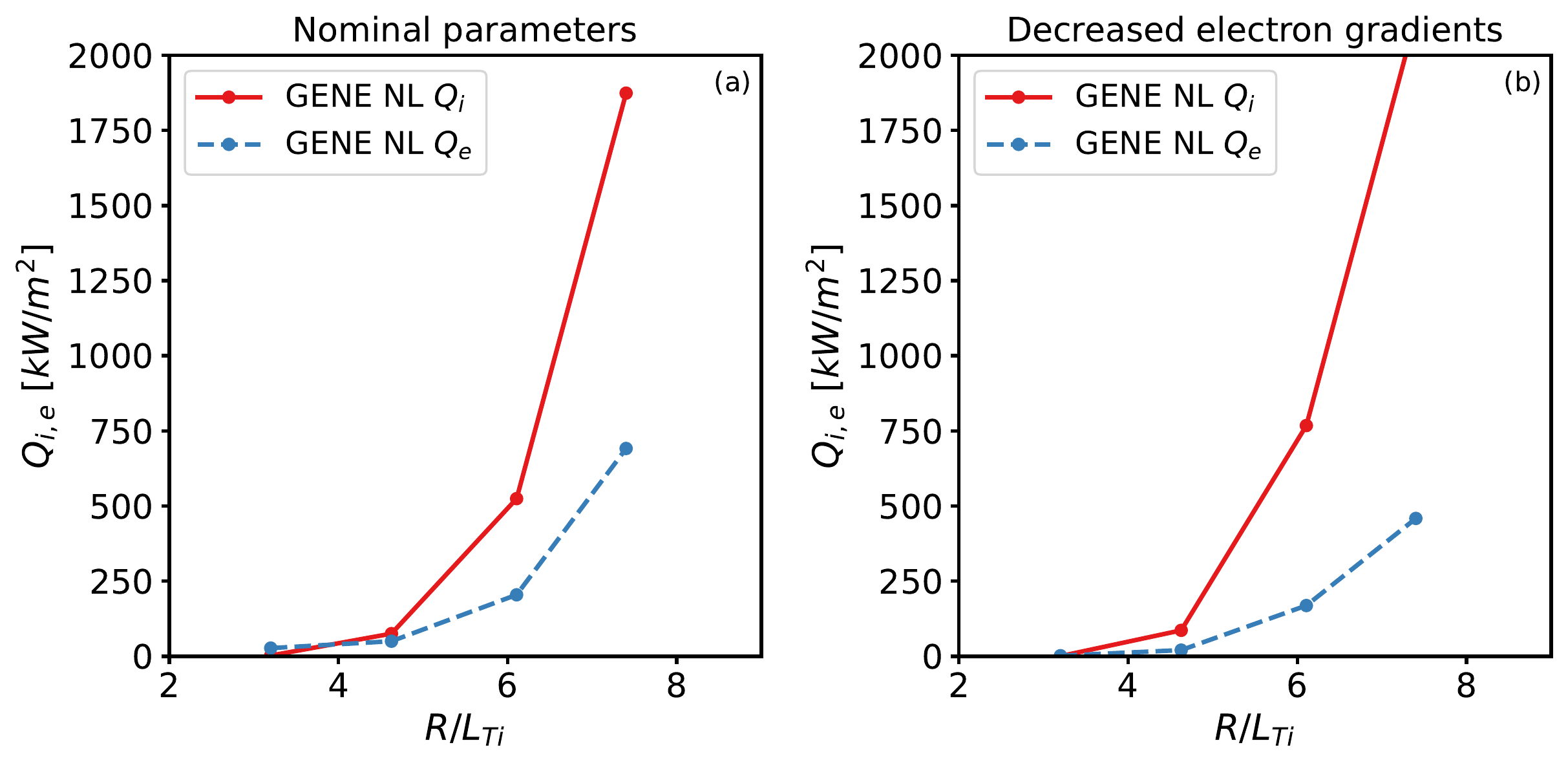}
	\caption{Ion and electron heat flux from nonlinear {\gene} {\rlti} scans, with non-{\rlti} input parameters corresponding to $\rho_N=0.5$ (see table~\ref{tab:physicsparams}). Panel (a) corresponds to nominal parameters, while panel (b) has {\rlte} and {\rlne} reduced by 30\% and is in a more dominant ITG regime with $Q_i>Q_e$ throughout the scan. } 
	\label{fig:nonlinearflux}
\end{figure}

The ion and electron heat fluxes from the nonlinear simulations are shown in figure~\ref{fig:nonlinearflux}. While the simulation uses dimensionless quantities, the output fluxes in the figure are converted to physical SI units using the ITER reference dimensional quantities at $\rho_N=0.5$ with $T_{\rm ref}=T_{\rm e}$ and $n_{\rm ref}=n_{\rm e}$, taken from the JINTRAC-QuaLiKiz simulation. For physical context, the ITER heat flux at $\rho_N=0.5$ from the integrated modelling simulation was $Q_{\rm i}\approx 200 kW/m^2$ and $Q_{\rm e}\approx 100 kW/m^2$, indicating that the stationary state is found in the vicinity of the turbulence thresholds. 

As elucidated in the next section, the nonlinear simulation heat fluxes, as well as more detailed information such as the wavenumber spectrum of the saturated electrostatic potential $|\delta\phi(k_y)|^2$, are used to constrain the quasilinear saturation rule.

\subsubsection{Saturation rule}
\label{sec:saturation}

The applied saturation rule is a variation of the QuaLiKiz saturation model~\cite{bourdelle:2007,bourdelle:2015,citrin:2017}. The differences are that we calibrate the model with specific nonlinear simulations carried out in our relatively narrow parameter space of interest, and also include a zonal flow mode suppression rule, needed to capture the correct $|\delta\phi_k|^2$ spectral shape from the nonlinear simulations, where $k$ denotes the binormal wavenumber $k_y$. We summarize the model briefly. 

Key quantities obtained from the linear simulations are mode growth rates $\gamma_k$, mode frequencies $\omega_k$, parallel mode structure $\phi_k(z)$ (with arbitrary normalization), and spatially averaged (local to a flux surface) transport fluxes. In our case, the transport fluxes per wavenumber are the ion heat flux $\bar{Q}_{i,k}$, electron heat flux $\bar{Q}_{e,k}$, and electron particle flux $\bar{\Gamma}_{e,k}$. The fluxes, which in the linear simulation exponentially grow, are normalized to the (similarly growing) electrostatic potential $|\phi_k|^2$ to provide a constant value. The bar over the fluxes denotes the normalization and spatiotemporal averaging. The quasilinear fluxes are thus defined as the following mode summation. 
\begin{equation}
\label{eq:fluxes}
    [Q_{i}, Q_{e}, \Gamma_{e}]=\Sigma_k|\phi_{sat,k}|^2[\bar{Q}_{i,k}, \bar{Q}_{e,k}, \bar{\Gamma}_{e,k}]
\end{equation}
The $k_y$ spectrum in the summation corresponds to all unstable modes for a given parameter configuration (variables listed in Table~\ref{tab:physicsparams}). $|\phi_{sat,k}|^2$ is the modelled saturated potential, calibrated to match the nonlinear simulations. The saturated potential model is: 
\begin{equation}
\label{eq:satrule}
|\phi_{sat,k}|^2 = C_{NL}\frac{\gamma_k}{k_\perp^2}|_{max}\frac{1}{k_{max}}\left(\frac{k_y}{k_{max}}\right)^\alpha
\end{equation}
$k_{max}$ is the modelled peak of the $|\phi_{sat,k}|^2$ spectrum, calculated as the $k_y$ corresponding to the maximum $\frac{\gamma_{k,eff}}{k^2_\perp}$ in the instability spectrum, where:
\begin{equation}
\begin{aligned} 
\gamma_{k,eff}&\equiv\mathrm{max}(0,\gamma_k-\omega_{ZF}) \\ 
\omega_{ZF}&\equiv\beta\mathrm{max}(\gamma_k),~\mathrm{for}~k_y<0.5
\end{aligned}
\end{equation}
$\omega_{ZF}$ is motivated by the effective $E{\times}B$ velocity shear rate due to zonal flows~\cite{staebler:2007}, and treated here as being proportional to the linear growth rate at the transport driving wavenumbers. This model was necessary for obtaining the saturated potential $k_{y,max}$ in agreement with nonlinear simulations. $\alpha$ and $\beta$ are free parameters calibrated to the nonlinear simulations, such that the saturation rule predicted $|\phi_{k}|^2$ spectra match the $|\phi_{k}|^2$ spectra from the nonlinear simulations. This procedure was carried out by eye, feasible due to the small number of cases. However, for fitting a saturation rule to a more extensive set of nonlinear simulations, needed for a more general quasilinear model across parameter space, a more principled weighted optimization routine must be developed. $\alpha=-2.25$ for $k_y>k_{max}$, $\alpha=1.5$ for $k_y<k_{max}$, and $\beta=0.5$. $\frac{\gamma_k}{k_\perp^2}|_{max}$ is calculated at $k_{max}$. Note that this means that $\frac{\gamma_k}{k_\perp^2}|_{max}$ is constant in the mode summation in equation~\ref{eq:fluxes}, meaning that effectively only an instability at a single wavenumber ($k_{max}$) contributes to the saturated potential for each spectrum in configuration space. However, all instabilities still contribute to the total flux through the $k$-dependent transport fluxes, which is the second term in equation~\ref{eq:fluxes}. Also note that the $k_y<0.5$ cutoff of the effective zonal flow shear rate implies that any instability threshold shift is likely insignificant, since the linear instability spectra tend to peak in the $k_y\sim0.5$ region. $C_{NL}$ is a constant parameter, calibrated such that the model ion heat flux matches the nonlinear simulation at $\rho_N=0.5$ with nominal parameters apart from {\rlti}$=6.02$. Finally, $k_\perp^2$ is the square of the perpendicular wavenumber conceptually split as $k_\perp^2=k_{\perp,linear}^2+k_{\perp,nonlinear}^2$. {\gene} linear mode output provides $k_{\perp,linear}^2$, which depends on $k_y$, geometric metric coefficients, and spatial mode structure, based on a generalized form of Eq. 4 in Ref.~\cite{dannert:2005}. $k_{\perp,nonlinear}^2$ is a modelled nonlinear contribution, motivated by zonal flow shearing of radial mode structures, and is based on Eq. 5 of Ref.~\cite{citrin:2012}. 

\begin{figure*}[hbt]
	\centering
	\includegraphics[width=1.0\linewidth]{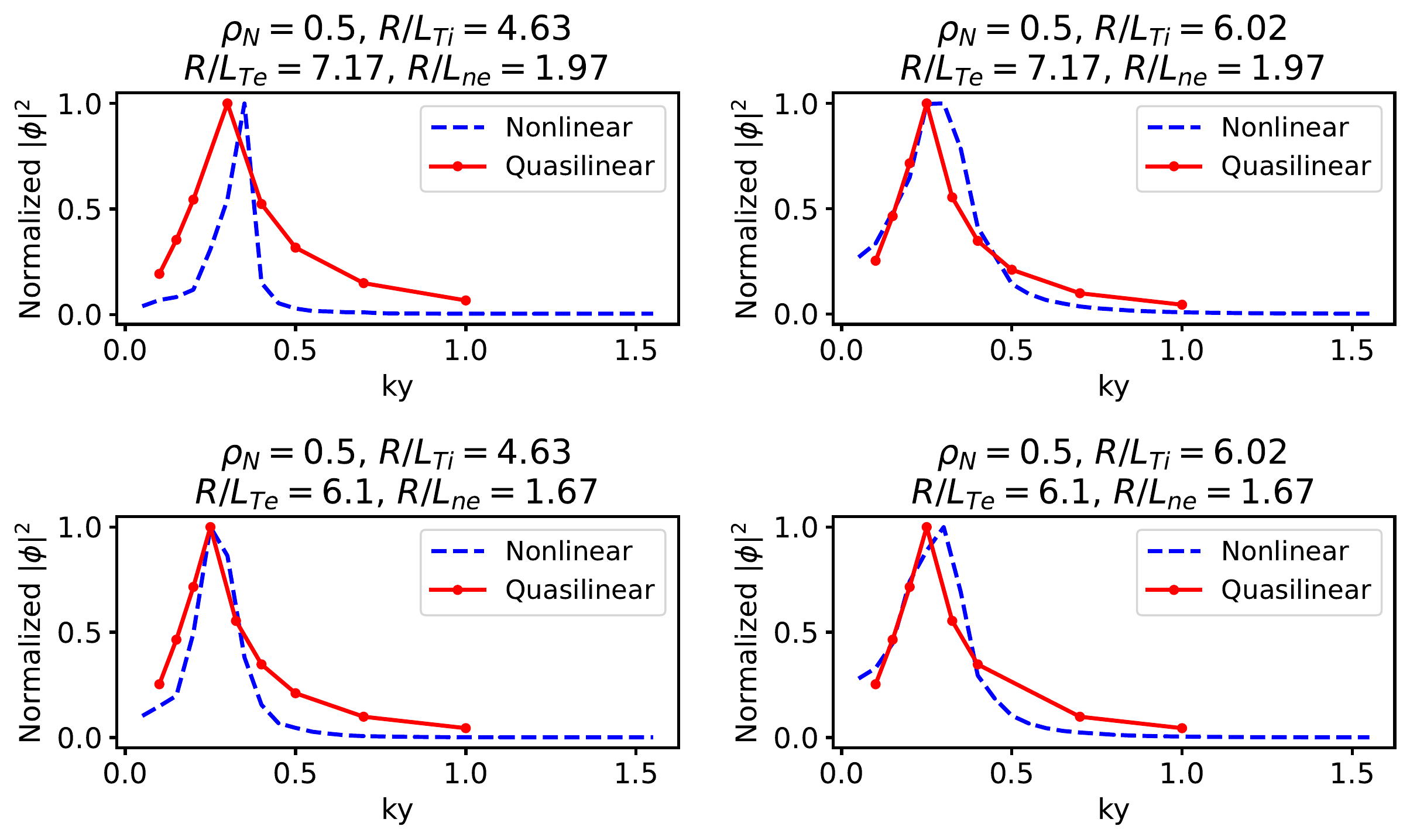}
	\caption{Normalized electrostatic potential spectra from a set of {\gene} nonlinear simulations (blue), compared to the modelled electrostatic potential spectra from the quasilinear saturation rule (red), for four of the nonlinear simulations carried out, corresponding to parameter configurations at $\rho_N=0.5$, at various {\rlti}, {\rlte}, and {\rlne}.} 
	\label{fig:calibrate_phi}
\end{figure*}

Validation of the saturation rule $|\phi_{k}|^2$ spectral model is shown in figure~\ref{fig:calibrate_phi}. $|\phi_{k}|^2$ spectra, averaged over $k_x$, are shown for four of the nonlinear simulations, and compared to the modelled $|\phi_k|^2$ in the saturation rule. All spectra are normalized to a maximum amplitude of unity, such that the spectral peaks and locations can be more easily compared. General agreement is observed for the spectral peak location ($k_{max}$) as well as the spectral shape. Exact correspondence between the spectral peaks cannot be expected due to the limited spectral resolution of the quasilinear model (nine wavenumbers). The worst agreement for the spectral shape is for the {\rlti}$=4.63$, {\rlte}$=7.17$, {\rlne}$=1.97$ case (top left panel), which is in a TEM regime. The narrow spectral shape in the nonlinear TEM simulation is reminiscent of toroidal mode condensation in TEM turbulence, as previously observed~\cite{citrin:2017b}. While the $\alpha$ and $\beta$ parameters in the saturation rule are similar in absolute value to the comparable saturation rules in Refs.~\cite{staebler:2007,casati:2009b}, the generality of the calibrated values cannot be claimed, since they would need to be compared to a wider variety of nonlinear simulations. Any generalization of this specific quasilinear model beyond the ITER baseline regime demands a generalization of the saturation rule calibration, likely introducing further parameter dependencies into the spectral model.

\begin{figure*}[hbt]
	\centering
	\includegraphics[width=1.0\linewidth]{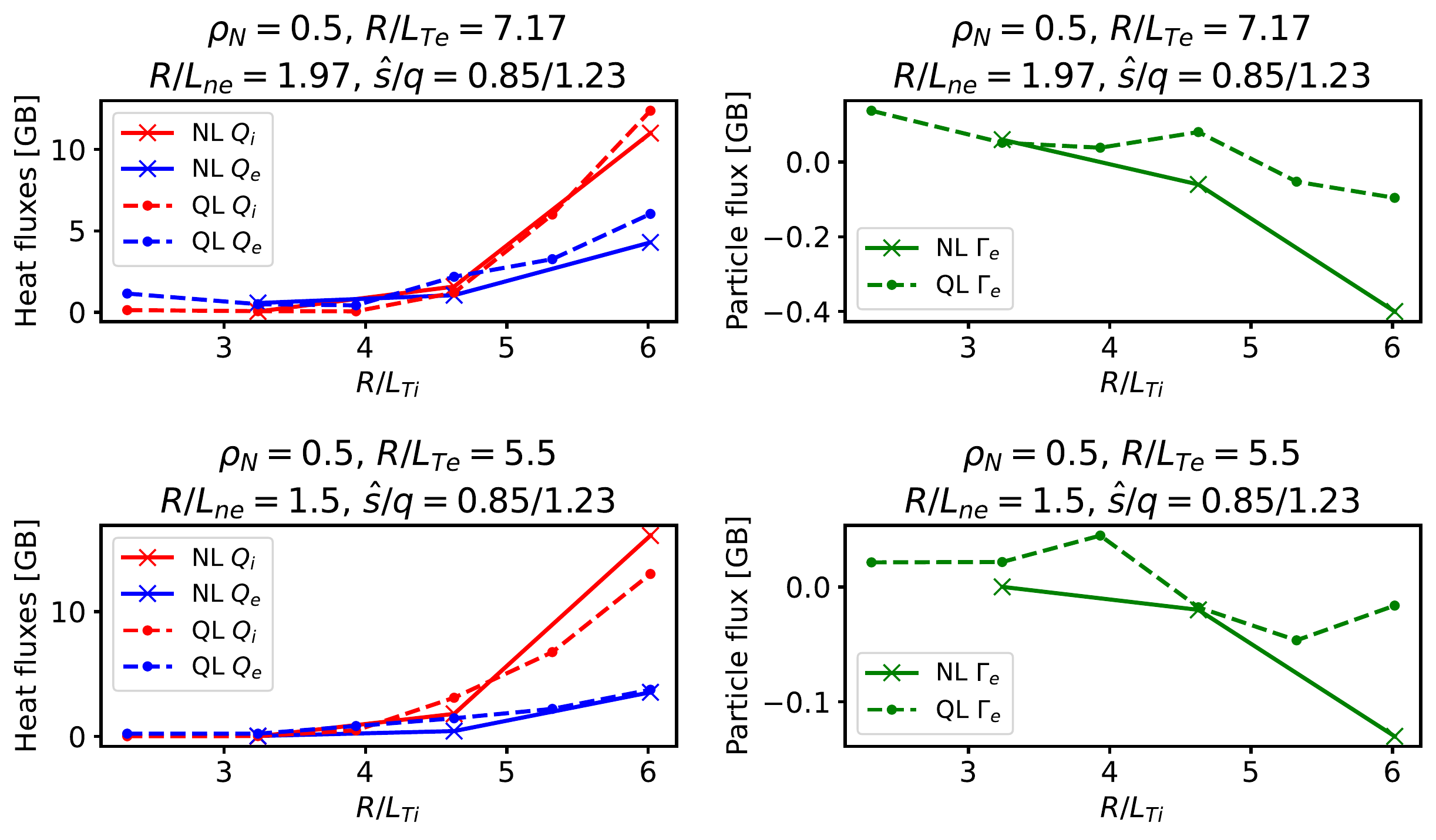}
	\caption{Comparison of ion heat fluxes ($Q_i$), electron heat fluxes ($Q_e$) and electron particle fluxes ($\Gamma_e$) between both sets of nonlinear {\gene} {\rlti} scans, and the quasilinear {\gene} model. All fluxes are in gyroBohm normalized units.} 
	\label{fig:calibrate_fluxes}
\end{figure*}

The comparison between the quasilinear and nonlinear fluxes is shown in figure~\ref{fig:calibrate_fluxes}, for all nonlinear simulations carried out in this study. Close correspondence between the heat fluxes is found, even for non-trival phenomena such as the TEM-ITG transition observed in the upper row with {\rlte}$=7.17$. Note that the flux correspondence is close even for the {\rlti}=$4.63$ case, which corresponded to the worst agreement in spectral shape, indicating robustness of the spectral shape model.  Regarding particle transport, non-trivial phenomena such as the transition from positive to negative flux is reproduced by the quasilinear model. Particularly at high {\rlti}, a deviation is observed in the particle fluxes, with the nonlinear fluxes systematically more negative. It is uncertain whether this is due to fundamental differences between nonlinear and quasilinear particle transport, as previously reported~\cite{baiocchi:2015b}, or the saturation rule. Regardless, at lower {\rlti} closer to the physical heat flux values, the deviations are more minor. The fluxes are gyroBohm normalized, which for the reference values taken for this study implies that for conversion to SI units the heat fluxes are multiplied by $Q_{GB}{\equiv}\frac{n_eT_e^{2.5}m_i^{0.5}}{q_e^2B^2R^2}$, and the particle fluxes by $\Gamma_{GB}\equiv\frac{T_e^{1.5}m_i^{0.5}}{q_e^2B^2R^2}$. GyroBohm normalization maintains the natural scalings of the local gyrokinetic system; the quasilinear training set and subsequent surrogate model NN are gyroBohm normalized, with denormalization to SI units occurring in the transport simulation itself. 

The saturation rule was also carried out for ITG modes and TEMs separately, by including all modes for the $k_{max}$ calculation but isolating ITG and TEM in the Eq.~\ref{eq:fluxes} summation. The training set was populated with these separate flux contributions, and the NN trained for ITG and TEM separately. In applications, the fluxes of the separate modes are added to the total flux. As discussed in Ref.~\cite{vandeplassche:2020}, this is necessary for cleanly capturing critical thresholds in the NN regression. 

\subsubsection{Data filtering and stable region supplementation}
\label{sec:filtering}
Additional post-processing was necessary for the final training set applied for NN model generation, due to the aforementioned challenges with convergence monitoring. All individual datapoints which were non-monotonic in $Q_i$ with increasing {\rlti} (for ITG) or non-monotonic in $Q_e$ with increasing {\rlte} (for TEM), were removed from the dataset. These points were assumed to arise due to non-convergence of low $k_y$ modes in that specific spectrum, leading to unphysical discontinuities in the gradient-flux response. By non-monotonic we mean lower in absolute $Q_i$ (for ITG) or $Q_e$ (for TEM) flux compared to the adjacent datapoint lower in the driving gradient scan, leading to that single datapoint to be removed. In addition, numerous stable points (zero fluxes) were not present in the dataset due to non-convergence of stable modes and thus inconclusive data. Zero flux data was supplemented by identifying datapoints where no data was received from the run generation pipeline, and checking whether any converged non-zero fluxes were present at lower {\rlti}, {\rlte}, or {\rlne}. If not, then that parameter configuration was identified as stable and the training set was supplemented by setting that datapoint as zero flux. If so, then due to the inconclusiveness, the datapoint was not included in the training set. Following filtering, approximately 2000 points were removed, corresponding to $\approx20\%$ of the quasilinear flux dataset. The total dataset size following both filtering and data imputation is 9449 entries per transport flux.

\section{Neural network model training and validation}
\label{sec:NNtraining}
The NN training follows similar methodology as the QuaLiKiz-neural-network (QLKNN), outlined in Refs.~\cite{vandeplassche:2020,ho:2021}. We summarize below. 

The model architecture consists of fully connected feed forward neural networks (FFNN), with three or four (depending on specific label) hidden layers of 128 nodes each, and a $tanh$ activation function for all nodes at each hidden layer. The input features are $\rho_N$, $\hat{s}$, $q$, {\rlti}, {\rlte} and {\rlne}. A total of six NNs were trained, each corresponding to a separate output label. In practice, training a single NN for each output label led to better regression accuracy compared to training multiple-output NNs. The labels correspond to two ``leading flux'' labels $Q_{i,ITG}$ (A) and $Q_{e,TEM}$ (B), and four ``division'' labels, $Q_{e,ITG}/Q_{i,ITG}$ (C), $\Gamma_{e,ITG}/Q_{i,ITG}$ (D), $Q_{i,TEM}/Q_{e,TEM}$ (E), and $\Gamma_{e,TEM}/Q_{e,TEM}$ (F). The leading flux networks (A) and (B) are characterized by sharp critical gradient threshold with respect to {\rlti} or {\rlte} respectively. Outputs of a leading flux and division networks are multiplied in post-processing to provide the full set of fluxes. For example, $Q_{e,ITG}$ arises from multiplying outputs of networks (A) and (C). This structure instills a physics-aware approach which ensures that all transport fluxes of a given mode class share identical critical gradient thresholds, important for accuracy in integrated modelling. See figure~6 in Ref.~\cite{vandeplassche:2020} for the consequences of applying a physics-unaware neural network model. All features and labels are normalized for training to mean $0$ and std $1$, using a normal distribution. 

For the division networks, only labels corresponding to unstable modes (with leading flux larger than zero) were provided in training. There, the optimization cost function was Mean Squared Error (MSE) with L2 regularization. However, for the leading flux networks themselves, both stable and unstable modes were included. The leading flux optimization cost function then included a custom provision in the loss term to capture the prior knowledge of sharp discontinuous critical gradient thresholds as follows:
\begin{equation}
\label{eq:costfunction}
    \left\{\begin{aligned}
    C_{penalty} &=C_{MSE}+C_{stab} \\
    C_{MSE}&=(y_{target}-y_{pred})^2,~~\text{if $y_{target}>0$} \\
         C_{stab}&=\lambda_{stab}y_{pred}^2,~~\text{if $y_{target}=0$, and $y_{pred}>0$}
    \end{aligned}\right.
\end{equation}
$C_{penalty}$ is the loss term per point. $C_{MSE}$ is the MSE penalty term, but only evaluated for unstable target fluxes. $C_{stab}$ is a penalty term only applied for stable flux targets, and when positive fluxes are erroneously predicted by the NN model. In post-processing, leading fluxes predicted negative by the model are then clipped to zero. This construct provides sharp critical gradient thresholds in spite of the L2 regularization. 

Training was carried out with Adam optimization with a learning rate of $0.001$, $\beta_1=0.9$, $\beta_2=0.999$, and a 80-10-10 split between training, validation, and test data. Early stopping criteria was applied on the validation set, with a patience of 100 epochs. L2 regularization hyperparameters, number of layers, and layer size, were optimized by hand based on visual inspection of model performance across randomized 1D scans of {\rlti}, {\rlte}, and {\rlne} throughout the data set. A summary of model architecture and hyperparameters is shown in table~\ref{tab:trainingchoices}. The Keras library was used for training~\cite{chollet2015keras}.

\begin{table*}
    \caption{Summary table of network architecture and hyperparameter values for the 6 separate NNs trained in this work. $\lambda_{stab}$ is defined in equation~\ref{eq:costfunction} describing the custom cost function applied for the ``leading flux'' NNs (A) and (B).}
    \centering
      \begin{tabular}{c|c|c|c|c|c|c|c}
	   NN label & Number of    & Activation & Nodes per    &  L2 & Batch & Patience & $\lambda_{stab}$ \\
        & hidden layers&  function  & hidden layer &    & size & \\
        \hline
	$Q_{i,ITG}$ (A) & 3  & $\mathrm{tanh}$ & 128 & $1e^{-4}$  & 32 & 100 & 0.2\\
	 $Q_{e,TEM}$ (B) & 3  & $\mathrm{tanh}$ & 128 & $2e^{-4}$  & 32 & 100 & 0.2\\
	 $Q_{e,ITG}/Q_{i,ITG}$ (C) & 3  & $\mathrm{tanh}$ & 128 & $5e^{-4}$  & 32 & 100 & N/A\\  
	 $\Gamma_{e,ITG}/Q_{i,ITG}$ (D) & 4  & $\mathrm{tanh}$ & 128 & $3e^{-4}$  & 32 & 100 & N/A\\  
	 $Q_{i,TEM}/Q_{e,TEM}$ (E) & 3  & $\mathrm{tanh}$ & 128 & $1e^{-4}$  & 32 & 100 & N/A\\  
	 $\Gamma_{e,TEM}/Q_{e,TEM}$ (F) & 3  & $\mathrm{tanh}$ & 128 & $1e^{-4}$  & 32 & 100 & N/A
	\end{tabular}
	\label{tab:trainingchoices}
	\normalsize
\end{table*}

\begin{figure*}[hbt]
	\centering
	\includegraphics[width=1.0\linewidth]{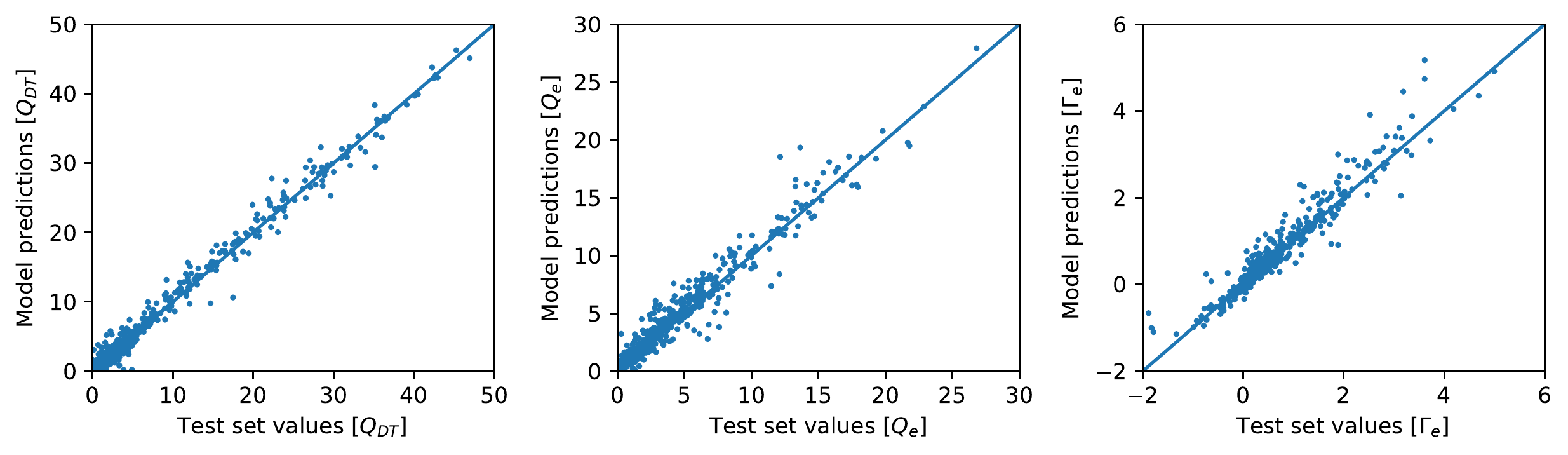}
	\caption{Comparison between the neural network predictions (y-axis) and test set values (x-axis) drawn from the {\gene} quasilinear flux database, for ion heat flux (left panel), electron heat flux (centre panel), and electron particle flux (right panel), all normalized to gyroBohm units.} 
	\label{fig:NN_testset_scatterplots}
\end{figure*}

\begin{figure*}[hbt]
	\centering
	\includegraphics[width=1.0\linewidth]{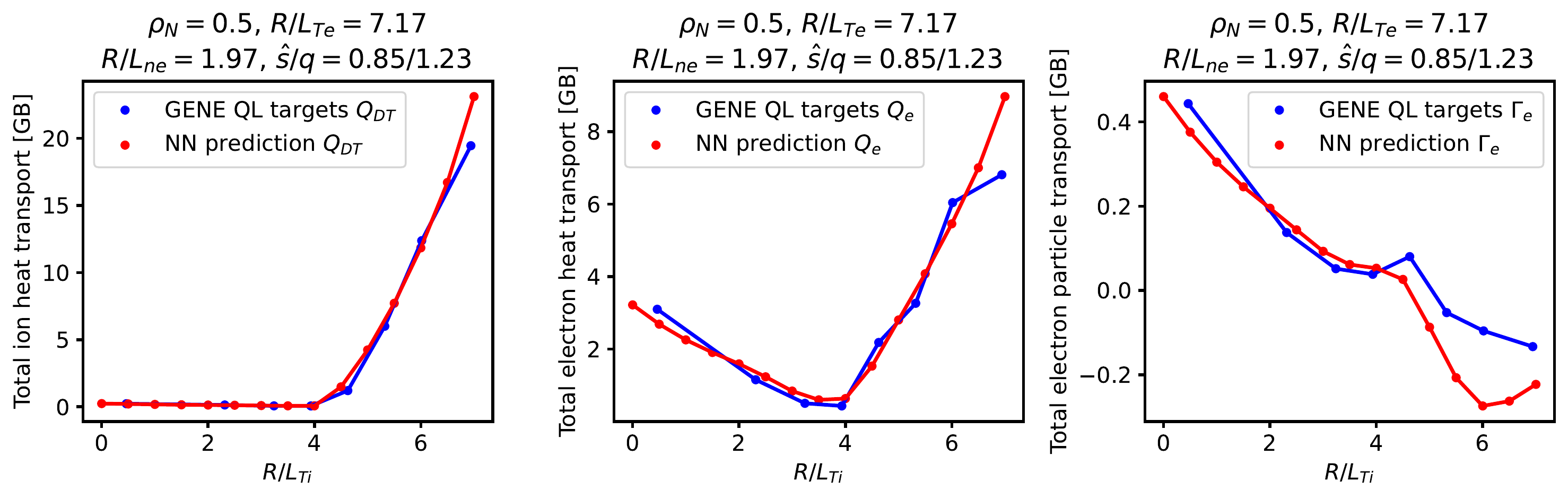}
	\caption{Comparison between the neural network model (red) and target labels from the {\gene} quasilinear flux database (blue) for ion heat flux, electron heat flux, and electron particle flux, all normalized to gyroBohm units. No distinction is made in the {\gene} data shown here, between the training, validation, and test sets used for the neural network training itself.} 
	\label{fig:NN_performance_1D}
\end{figure*}

Demonstration of model performance is shown in figure~\ref{fig:NN_testset_scatterplots}, comprising of scatter plots comparing test set values with model predictions for ion heat flux, electron heat flux, and electron particle transport. Total fluxes are shown, combining the ITG and TEM fluxes. The model predictions thus arise from combining several NNs. Following the labelling in table~\ref{tab:trainingchoices}: $Q_{DT}=A+B{\cdot}E$, $Q_e=B+A{\cdot}C$, and $\Gamma_e=A{\cdot}D+B{\cdot}F$. Model performance in a 1D  logarithmic ion temperature gradient scan is shown in figure~\ref{fig:NN_performance_1D}, for the same case at $\rho_N=0.5$ as in the top row of figure~\ref{fig:calibrate_fluxes}. Reproduction of salient structural features of the ion temperature gradient to turbulent flux relation is evident. This includes sharp ITG critical thresholds, and non-monotonic electron heat flux due to the TEM to ITG transition. While the zero-flux crossing is reproduced well, the NN particle flux prediction deviates from the {\gene} labels at higher {\rlti}. This is due to inherent discontinuous structures in the data caused by either mode convergence issues or saturation rule $k_{max}$ discontinuities, which has more of an impact on the smoothness of the particle flux data, since particle flux per mode can be positive or negative, potentially leading to a loss of local monotonicity. The NN model regularization provides smoothing in the multidimensional space, reducing the impact of these discontinuities, which also contribute to the deviations evident in figure~\ref{fig:NN_testset_scatterplots}. 

\begin{figure}[hbt]
	\centering
	\includegraphics[width=1.0\linewidth]{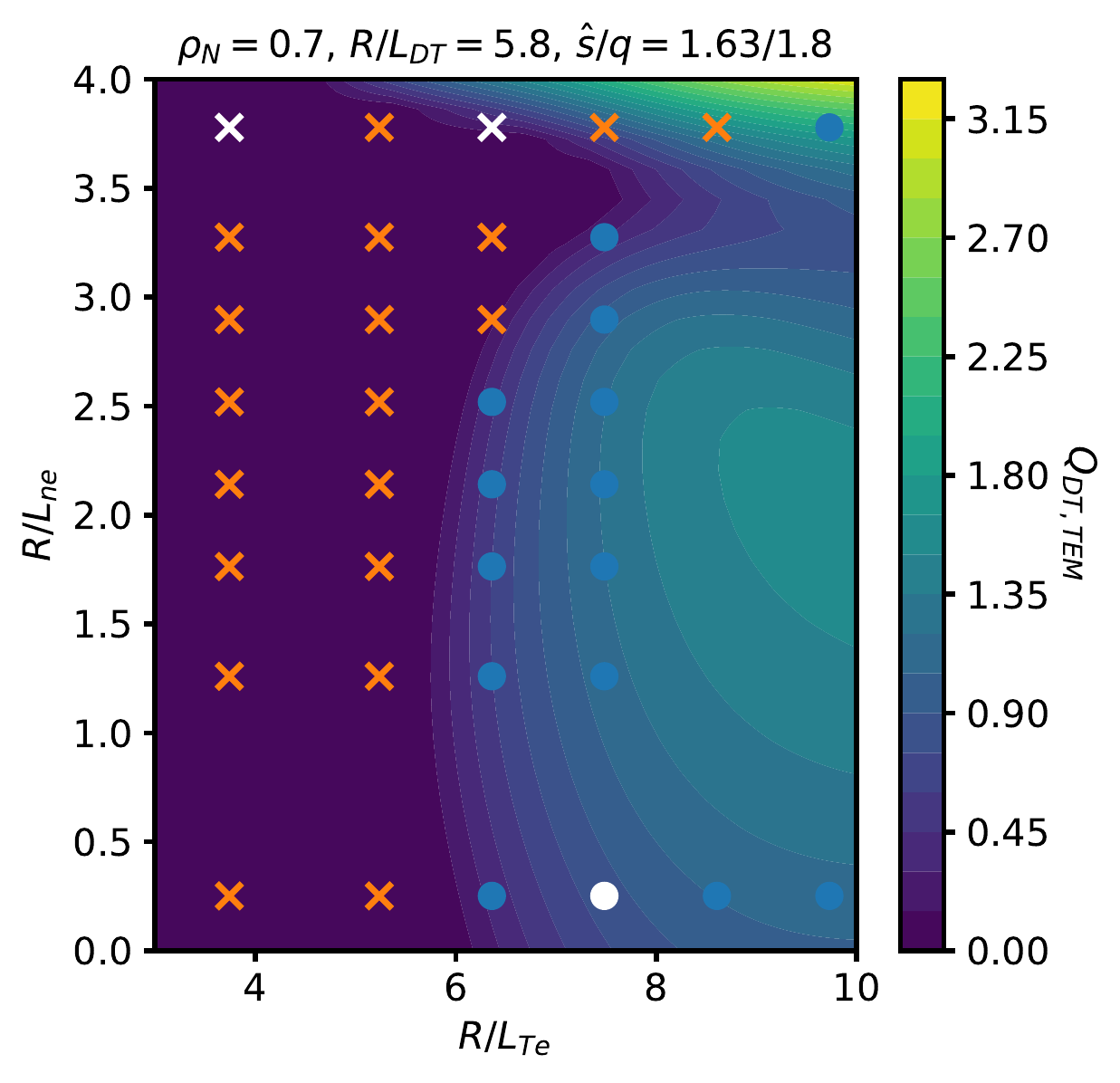}
	\caption{Comparison between NN model (contour plot) and the {\gene} quasilinear dataset (discrete points) for TEM electron heat flux (gyroBohm normalized), varying both {\rlte} and {\rlne}, the two main driving gradients of TEM instabilities. White points belong to the test set, and the rest to the training and validation sets. The information from the points corresponding to the {\gene} database is reduced to whether the point is stable (crosses), or unstable (circles). This is to more easily discern the agreement with the critical threshold behaviour.}
	\label{fig:NN_performance_2D}
\end{figure}

Further insight is gained through a 2D parameter scan, shown in figure~\ref{fig:NN_performance_2D}, for TEM electron heat flux at $\rho_N=0.7$. The figure illustrates that in spite of complex 2D structure of the gradient-flux relationship, the key component of capturing the discontinuous critical thresholds is successfully carried out. Nevertheless, note that two false positives ($Q_{NN}>0$ while $Q_{dataset}=0$) are evident in the top row, where the structure is the most complex.

Performance metrics for the NN models are shown in table~\ref{tab:metrics}, comparing NN output with the full original dataset. The average absolute error is compared with the average values of each flux. This metric was chosen, as opposed to relative mean square error, to avoid overly weighting the impact of errors near critical threshold where absolute error matters less, due to high transport stiffness, and for zero particle flux (which is not necessarily tied to the critical threshold). The relatively larger error for the particle flux likely arises from both the inherent increased nonlinearity of the particle flux dataset, as well as the propagation of errors in the multiple networks used in the calculation (particle flux is never a leading flux). A degree of error is also always expected due to the regularization, which smooths over irregularities in the data. In addition, the $R^2$ scores for the test set is shown, where a similar pattern emerges showing larger errors for the electron heat and particle fluxes.

Finally, a metric for ``mislabeling''of the output is provided. Namely, the percentage out of all stable (zero) fluxes in the dataset, where the NNs predict non-zero fluxes (spurious flux ratio), and the percentage out of all unstable fluxes where the NN predicts stability (spurious stable ratio). As with the absolute error metrics, a degree of mislabelling can be expected due to the regularization and data irregularites. The average flux value of the spurious flux points is 0.39 in gyroBohm units, which is low, signifying that these are likely mostly close to the critical thresholds in practice. 

\begin{table*}
    \caption{Performance metrics for the {\gene} quasilinear NN model, compared to the full original dataset. The first three columns correspond to the average absolute error divided by the average values for all unstable ion heat flux, electron heat flux, and particle fluxes. The subsequent three columns correspond to the $R^2$ score for the unstable points in the test dataset. The values of the training dataset are similar and omitted for brevity. The last two columns signify the percentage of potential mislabeling, as in the percentage of NN predicted non-zero fluxes which were stable in the dataset, and vice versa. Spurious flux (false positive) is defined as cases where $Q_{NN}>0$ and $Q_{dataset}=0$. Spurious stable (false negative) is defined as cases where $Q_{NN}=0$ and $Q_{dataset}>0$.}
    \centering
      \begin{tabular}{c|c|c|c|c|c|c|c}
	  &  &  & & & & Spurious flux ratio & Spurious stable ratio \\
     	  $\frac{\langle|\delta Q_{DT}|\rangle}{\langle Q_{DT}\rangle}$     & $\frac{\langle|\delta Q_{e}|\rangle}{\langle Q_{e}\rangle}$ & $\frac{\langle|\delta \Gamma_e|\rangle}{\langle \Gamma_e\rangle}$ & $R^2(Q_{DT})$ & $R^2(Q_{e})$ & $R^2(\Gamma_{e})$ & (False positive ratio) & (False negative ratio) \\
        \hline
	    10\% & 16\%  & 29\% & 0.9896 & 0.9548 & 0.9196 & 10\%  & 7\%  \\
	\end{tabular}
	\label{tab:metrics}
	\normalsize
\end{table*}

In summary, the trained models are deemed a satisfactory representation of the {\gene}-QL dataset. The model was then coupled to an integrated modelling suite for ITER baseline scenario simulation.

\section{Application within integrated modelling}
\label{sec:integratedmodelling}
The {\gene} quasilinear neural network model ({\geneqlnn}) was coupled to the JINTRAC integrated modelling framework, as a drop-in replacement for standard quasilinear models such as QuaLiKiz and TGLF. Inputs from JINTRAC to {\geneqlnn} are the NN input features $\rho_N$, {\rlti}, {\rlte}, {\rlne}, $q$, and $\hat{s}$. Outputs from {\geneqlnn} to JINTRAC are the ion and electron heat fluxes, and electron particle flux, arising from combinations of the 6 separate NNs as described in section~\ref{sec:NNtraining}. The ITER baseline case ($B_T=5.3T, I_p=15MA$) shown in figure~\ref{fig:manticaITER} was rerun with {\geneqlnn}, with the same simulation settings, reviewed here briefly. Predicted plasma processes and respective quantities were poloidal flux diffusion (plasma current), heat transport (temperatures), and particle transport (density). The JINTRAC 1D Partial Differential Equation (PDE) radial resolution was 71 grid points. {\geneqlnn} was called on a downscaled grid of 25 points, with transport coefficients interpolated onto the full grid following each call. The timestep was adaptive, with a maximum set at 3.3~ms. Due to the predictor-corrector method employed, {\geneqlnn} was called twice per timestep, for each point on the downscaled grid. Grad-Shafranov plasma equilibrium was self-consistently calculated throughout with the ESCO code~\cite{cenacchi:1988}. The impurity content and radial profiles were set as constant in time, from the final stationary phase of the previous JINTRAC-QuaLiKiz simulation. However, impurity charge state equilibrium was self-consistently calculated with the evolving $T_e$ and $n_e$. 33MW of NBI heating was self-consistently calculated with the PENCIL code~\cite{challis:1989}. 20MW of Electron Cyclotron Resonance Heating (ECRH) was prescribed as a Gaussian with a peak location at $\rho_N=0.4$, and a FWHM of $\rho_N=0.13$. Neoclassical transport was calculated with the NCLASS code~\cite{houlberg:1997}. Pellet fuelling was simulated with a simplified discrete pellet model without drifts~\cite{garzotti:1997}. Line radiation from impurities was self-consistently calculated. The main ion (D and T) particle transport was set as the output of the NN electron particle flux, in gyroBohm units. This assumption is necessary since only electron particle fluxes were included in the training set, due to the pragmatic decision to maintain fixed impurity gradients. Alpha particle heating was calculated with a simple $T_e$-dependent parameterization that calculates the fractional energy deposition on ions and electrons, and assumes equivalence between the heating deposition and fusion rate profiles~\cite{mikkelsen:1983}. The model boundary condition was taken as the EPED-consistent pedestal top at $\rho_N=0.93$. In practice, the last grid point of model evaluation was at $\rho_N=0.8936$ (within the training limits), and a constant extrapolation of transport coefficients was assumed up until the internal boundary condition location at the pedestal top. The process in which a EPED-consistent pedestal is calculated for these ITER conditions is detailed in Ref.~\cite{koechl:2018}. We set the pedestal height obtained in Ref.~\cite{koechl:2018} as an internal boundary condition at $\rho_N=0.93$ and do not modify the pedestal self-consistently with the core profile evolution, which itself can modify the pedestal calculations, e.g. through modifications in plasma $\beta$~\cite{snyder:2011}. From the initial condition, the simulations last 20 seconds of plasma evolution, which is several energy confinement times and sufficient to reach a quasi-stationary state. With {\geneqlnn}, the simulation required 60 CPU minutes on a single core (Intel(R) Xeon(R) Processor E5-2665 @ 2.40GHz). {\geneqlnn} was not the bottleneck in the simulation. {\geneqlnn} model inference times is on the order of 1ms per PDE timestep ($\Delta t\sim3~ms$ for this case), and thus realtime capable for ITER. For comparison, the JINTRAC-QuaLiKiz simulation required $\approx500~CPUh$. If quasilinear {\gene} is directly applied as the turbulent transport model within integrated modelling, then the estimated required CPU time for this simulation would be 1 million CPUh. This time is more than was required for the training set generation ($\approx0.25$ million CPUh). It is striking that the extent of surrogate model speedup enables compute time to be saved following a single use.

\begin{figure*}[hbt]
	\centering
	\includegraphics[width=1.0\linewidth]{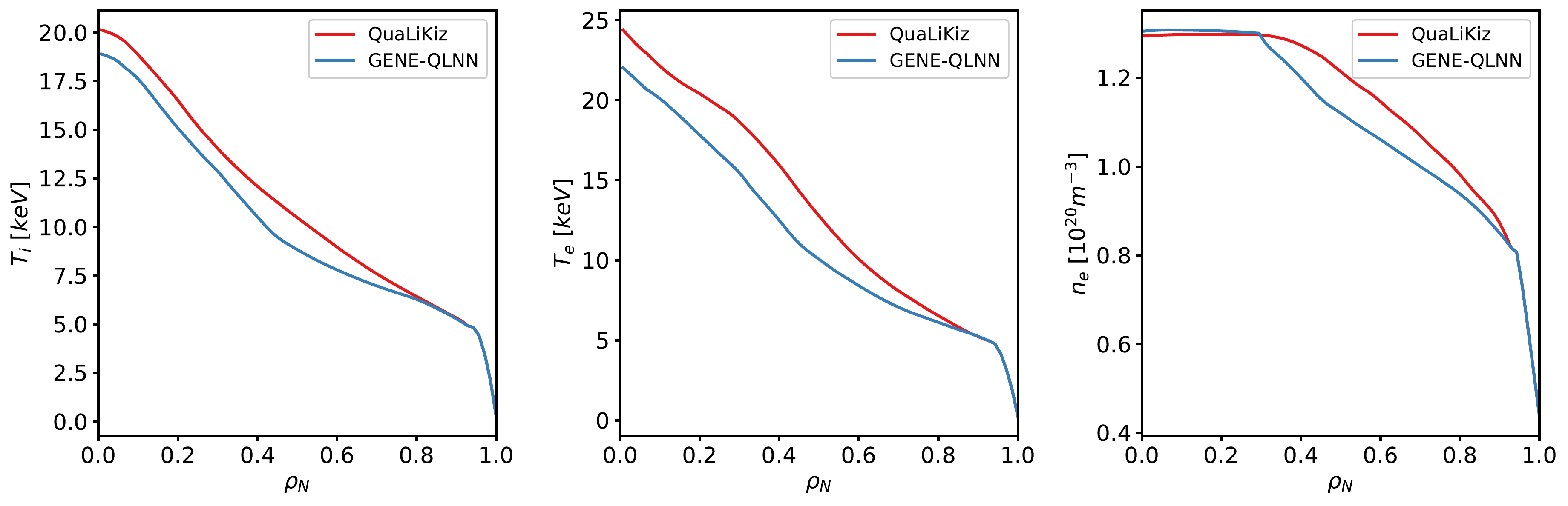}
	\caption{Comparison of JINTRAC-QuaLiKiz and JINTRAC-[{\geneqlnn}] multi-physics simulations of the ITER baseline scenario with $I_p=15MA$, for ion temperature (left panel), electron temperature (center panel), and electron density (right panel). The core boundary condition is taken at normalized toroidal flux coordinate $\rho_N=0.92$. The plots correspond to time-averages over the final 800~ms of the respective simulations (each lasting 20 plasma seconds), during quasi-stationary state.} 
	\label{fig:jetto_optimal}
\end{figure*}
\begin{figure}[hbt]
	\centering
	\includegraphics[width=0.75\linewidth]{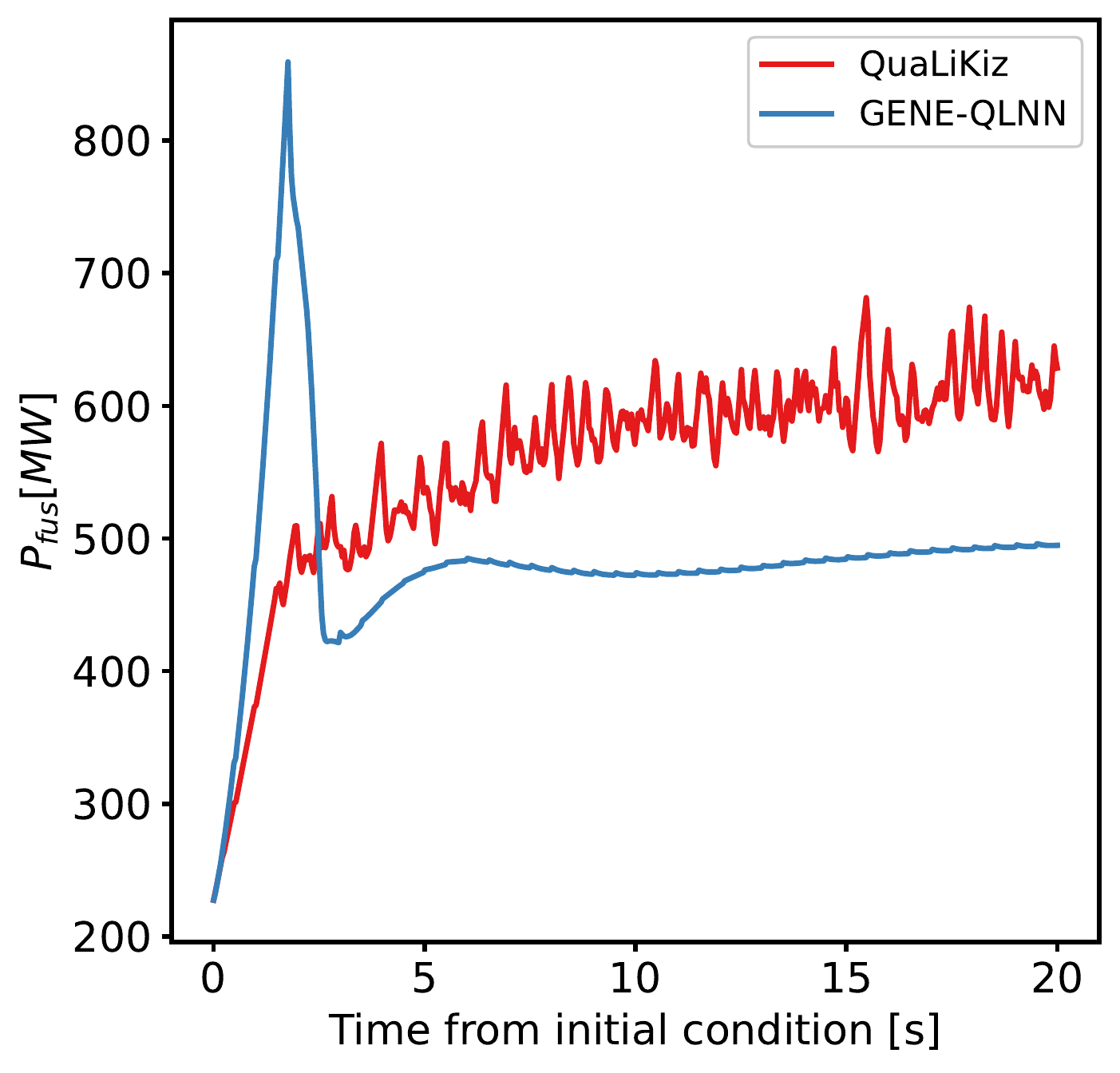}
	\caption{Fusion power output for JINTRAC-QuaLiKiz and JINTRAC-[{\geneqlnn}] ITER baseline simulations shown in figure~\ref{fig:jetto_optimal}. The QuaLiKiz timetrace has continuous jitter due to inherent fluctuations in the gradient-flux relationship over time, due to QuaLiKiz mode convergence issues, i.e strong sensitivity to input parameters of converging / not-converging to eigenvalue solutions near instability thresholds. The initial transient in the GENE-QLNN simulation is due to an initial density spike near the internal boundary condition ($\rho_N=0.93$) due to low turbulence at the initial condition, and the pellet particle source. The qualitatively difference initial trajectories between QuaLiKiz and GENE-QLNN may indicate that GENE-QLNN was transiently extrapolating outside its training set, even if the final stationary state (the goal of this use-case) is well within the training set.} 
	\label{fig:ITERpowertrace}
\end{figure}

The results are shown in figure~\ref{fig:jetto_optimal}. The JINTRAC-QuaLiKiz simulation from Ref.~\cite{mantica:2020} is compared with the JINTRAC-[{\geneqlnn}] simulation, for ion temperature, electron temperature, and electron density predictions at stationary state. Averaging over the last 1~s of the simulation, the {\geneqlnn} predictions correspond to approximately 18\% lower plasma confinement than QuaLiKiz, with $327~MJ$ stored thermal energy as opposed to $386~MJ$. This corresponds to $P_{fus}\approx495~MW$ (power from DT fusion reactions) for JINTRAC-[{\geneqlnn}], as opposed to $P_{fus}\approx619~MW$ for JINTRAC-QuaLiKiz. See figure~\ref{fig:ITERpowertrace} for the fusion power time traces. These power values correspond to fusion gain of $Q=9.36$ and $Q=11.7$ respectively. The ITER baseline goal is $Q=10$. The QuaLiKiz simulations were optimistic, and likely the lower confinement in the {\geneqlnn} simulations is due to the more correct treatment of TEMs, which are more strongly driven in {\gene} compared to the QuaLiKiz version used here~\cite{stephens:2021b}. However, the {\geneqlnn} results are still consistent with ITER goals. Nevertheless there are multiple sources of uncertainty in the projections, with identifiable trends in the modifications they would induce. The JINTRAC-[{\geneqlnn}] simulations were conservative by not incorporating the stabilizing impact of ${E \times B}$ shear or electromagnetic effects. Furthermore, the average $T_i/T_e$ is higher in the GENE-QLNN simulation by $\approx6\%$ compared to the original QuaLiKiz simulation; this would increase the ITG critical gradient thresholds in the GENE-QLNN simulation if $T_i/T_e$ is included as a GENE-QLNN input. However, the reduced core confinement predicted in the GENE-QLNN simulations as they stand, may also lead to a reduced pedestal height, due to a lower total plasma $\beta$~\cite{snyder:2011}. While recent EPED scans in the ITER baseline regime have shown small variation of the pedestal height with plasma-$\beta$~\cite{howard:private}, future work should nevertheless focus on self-consistent core-pedestal coupling, as in Ref.~\cite{meneghini:2020}. Additional sources of uncertainty are the impurity content, the transport coefficients in the central core region ($\rho_N<0.3$), which are prescribed, as well as the pedestal boundary condition height and width. A more detailed UQ study is out of the scope of this paper. A consequence of the central core region transport coefficient prescription is the observed density flattening. This arises due to a combination of a lack of turbulent transport (which provides an inward convective term leading to density peaking), a lack of central particle source (the pellet source is limited to the outer region), and the small prescribed background particle diffusivity (background $D_e=0.1~m^2/s$ in this case). The combination of residual diffusivity and lack of particle source leads to the flattening.

The {\geneqlnn} model applied here was trained with the full dataset from both Phase 1 and Phase 2 of run generation. At each radial location, the final {\rlti}, {\rlte}, {\rlne}, $\hat{s}$ and $q$ values do not extend beyond the training set envelope. In Appendix B we outline what occurs when applying a {\geneqlnn} version only utilizing the Phase 1 data, where the subsequent simulation extrapolates beyond the training set. 

\section{Conclusions and outlook}
\label{sec:conclusions}

Linear gyrokinetic calculations, nonlinear simulations, a bespoke saturation rule, and neural network regression have been combined to generate a fast and accurate surrogate turbulent transport model applicable for ITER baseline scenario modelling. The neural network inference time is eight orders of magnitude faster than the original quasilinear gyrokinetic calculations. The model predicts a fusion gain of $Q\approx9$ for the ITER baseline scenario, in line with ITER targets. The prediction is conservative since it does not take into account either the stabilizing impact of background perpendicular velocity shear or electromagnetic stabilization of ITG turbulence, which may increase confinement. This work provides the as-yet highest fidelity extrapolation of ITER baseline core plasma confinement. Self-consistent core-pedestal coupling is necessary to further refine the extrapolation.

There is significant opportunity to generalize the approach to wider parameter sets, input features, and model physics content, and provide a more broadly applicable transport model both faster and more accurate than present-day quasilinear transport models such as QuaLiKiz and TGLF. However, such extensions demand further innovations in a number of aspects. We list these challenges below, as well as potential solutions, left for future work:

\begin{itemize}
    \item The {\gene}-NN training set is likely over-sampled. This ensured sufficient data for this specific demonstrative application. However, for wider parameter sets the computational burden of linear gyrokinetic runs can become restrictive. Therefore, Active Learning pipelines where only truly informative data-points are simulated must be developed, as recently demonstrated in the fusion turbulence surrogate modelling context~\cite{burr:2022}.
    \item A multi-fidelity approach should be adopted, whereby a database of nonlinear simulations is strategically scattered in parameter space to help constrain the saturation rules and/or the surrogate models themselves. Criteria for how to distribute these high-fidelity simulations must be established.
    \item Improvements in mode convergence algorithms should be developed within the gyrokinetic codes themselves, to facilitate large-scale automated run generation pipelines. This includes improved automated identification of mode stability, automated detection of numerical convergence issues based on expert criteria, and automated averaging over quasi-stationary states as opposed to strict mode convergence, for those cases where this occurs.
    \item Information on subdominant modes can be important for accurate model descriptions in certain regimes. Advances in eigenvalue solution methodology is thus needed, particularly on robustness for large parameter scans where preconditioner parameters may need to vary in an automated way throughout parameter space. Such a capability is a prerequisite to generalize this approach to stellarator turbulence, where multiple subdominant modes need to be included for an accurate quasilinear model~\cite{pueschel:2016}.
    \item The approach of building surrogate models for each mode type separately, helpful for clean critical threshold capturing, will need to be re-evaluated in a more general setting such as electromagnetic turbulence where additional mode types can arise. These include Kinetic Ballooning Modes (KBMs), MicroTearing Modes (MTMs), and energetic particle driven modes. More complex criteria will need to be developed to identify mode types in an automated way. 
    \item Robustness against flux discontinuities due to non-converged modes can be achieved by training surrogate models directly on the linear mode characteristics (i.e. growth rates, frequencies, quasilinear transport weights, perpendicular wavenumber) instead of the transport fluxes. The transport model is then evaluated by applying a saturation rule on the surrogate model outputs, which will be smooth. This has the additional advantage of allowing various saturation models to be tested in a modular fashion. 
    \item Variations of the supervised learning model architecture should be explored. The NN topology can be structured to better capture the salient features of the input-output structure, e.g. through a parameterized critical-gradient-model~\cite{horn:2022}. Additionally, instead of FFNNs, decision tree regression models like XGBoost can be investigated. 
\end{itemize}

A high-fidelity fast surrogate turbulence model based directly on quasilinear and nonlinear gyrokinetics would provide significant advances in tokamak scenario optimization and control-oriented applications. Ultimately, the goal is to train innovative and general controllers directly from simulation frameworks~\cite{degrave:2022}. For enabling design of multivariate controllers, the simulation frameworks demand fast and accurate multi-physics simulation capability. The model and methodology we have introduced fits into this approach.


\section*{Data availability statement}
The {\gene} linear instability and quasilinear flux datasets calculated in this study have been published in an open source repository~\cite{citrin:dataset2023}. The NN models developed in this study are available upon reasonable request from the authors.

\section*{APPENDIX A: Justification of bundled ion assumption} 

\begin{figure}[hbt]
	\centering
	\includegraphics[width=1.0\linewidth]{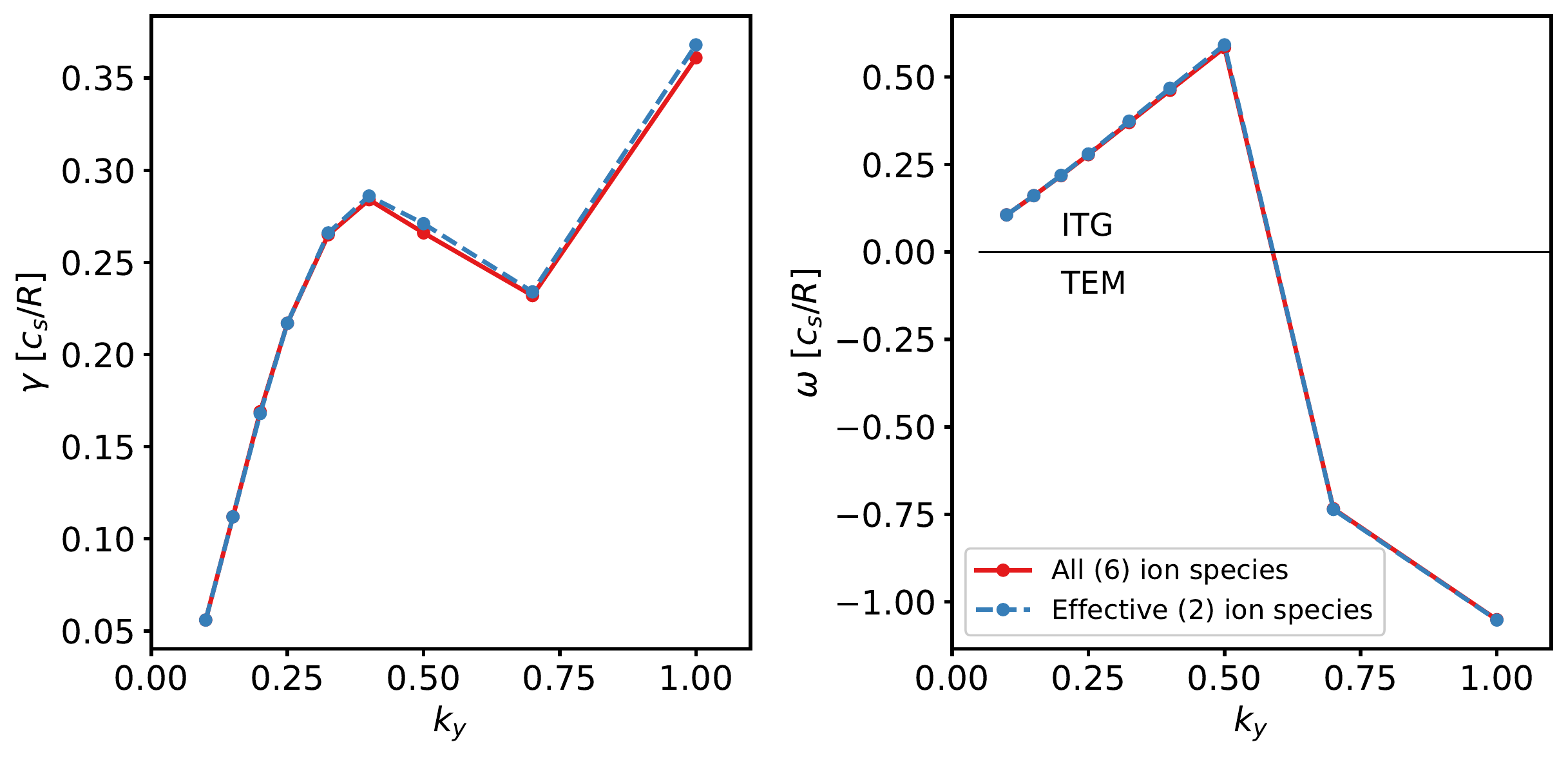}
	\caption{Comparison of linear {\gene} simulations corresponding to the nominal parameters at $\rho_N=0.6$. The resultant growth rates (left panel) and frequencies (right panel) are compared for input parameters containing all 6 ion species, and for 2 effective bundled ion species.}
	\label{fig:impuritycomp}
\end{figure}

As discussed in section~\ref{sec:trainingset}, the original ITER simulation contained 6 ion species: separate deuterium (D) and tritium (T), helium (He) fusion ash, beryllium (Be) and tungsten (W) arising from plasma-wall-interactions, and seeded neon (Ne) for heat exhaust control. To reduce the number of ion species and thus decrease the computational burden of the {\gene} runs, a bundled ion assumption was made. D and T were bundled into a single effective main ion species. All impurities (He, Be, W, Ne) were bundled into a single effective species with a charge state, density and density gradient chosen such that the main ion species maintains the same density and density gradient as D and T in the original 6-ion set, and that the effective charge $Z_\mathrm{eff}$ is maintained. The implication for this case was to set the effective impurity charge as $Z=7$. The effective impurity mass was set at $A=2Z$. Since the primary impact of impurities on ITG and trapped electron mode (TEM) instabilities -- which dominate this ITER case -- is through the impact on main ion density and density gradient, as well as collisionality, these choices minimize the impact of the bundled ion assumption on the resultant modes. This is illustrated in figure~\ref{fig:impuritycomp} for a representative case at normalized radius $\rho_N=0.6$, where agreement within 2\% is seen between 6-ion and effective 2-ion simulations. 

\section*{APPENDIX B: impact of only using Phase 1 data for surrogate model} 

\begin{figure*}[hbt]
	\centering
	\includegraphics[width=1.0\linewidth]{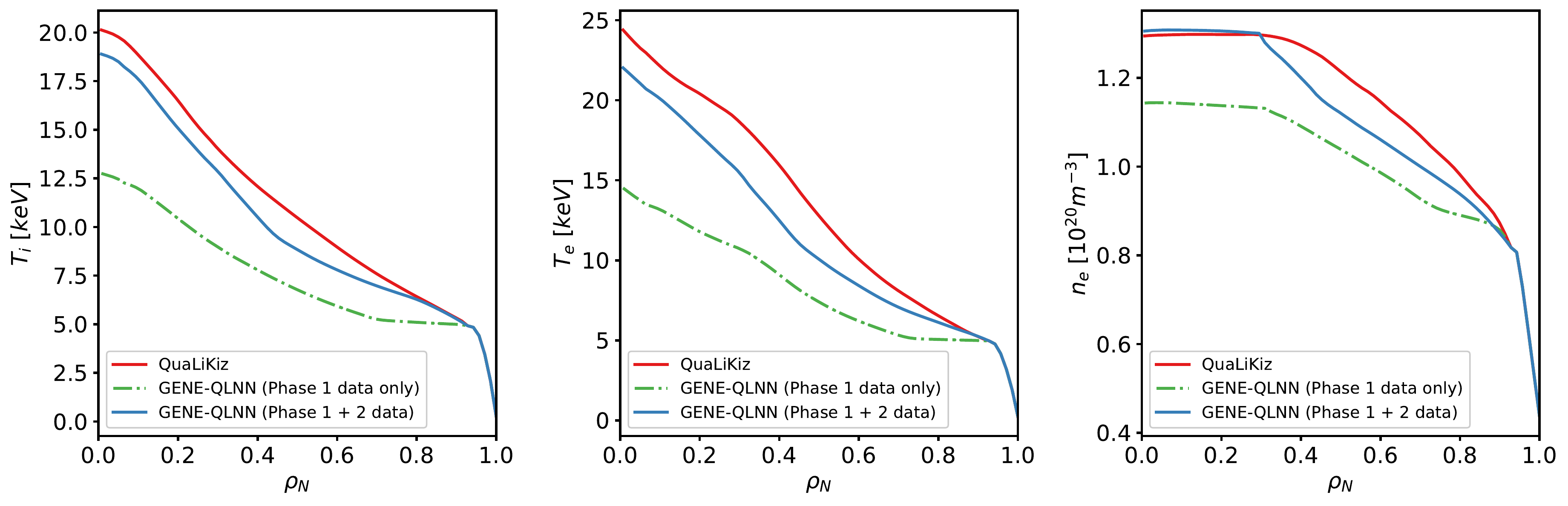}
	\caption{Comparison of JINTRAC-QuaLiKiz (red solid lines) and JINTRAC-[{\geneqlnn}] multi-physics simulations of the ITER baseline scenario with $I_p=15MA$, for ion temperature (left panel), electron temperature (center panel), and electron density (right panel). The case with {\geneqlnn} model trained only with the Phase 1 dataset (blue solid line) is compared with the complete case trained with both Phase 1+2 data (green dotted line). The Phase 1 case is extrapolating beyond its training set, evident through the temperature flattening in the $\rho_N\sim0.8$ region. The core boundary condition is taken at normalized toroidal flux coordinate $\rho_N=0.92$. The plots correspond to time-averages over the final 800~ms of the respective simulations, during quasi-stationary state.} 
	\label{fig:jetto_nonoptimal}
\end{figure*}

In figure~\ref{fig:jetto_nonoptimal} we show the results of a JINTRAC-[{\geneqlnn}] simulation of the ITER baseline scenario. We compare the simulation shown in figure~\ref{fig:jetto_optimal}, which had a surrogate model trained using the full Phase 1+2 datasets, with a model trained only with the Phase 1 dataset, which did not include the supplementary data points at low {\rlti}, {\rlte}, and {\rlne}. The simulation with the Phase 1 dataset model develops a distinct flattening of ion and electron temperatures in the $\rho_N\sim0.8$ region. This flattening corresponds to an extrapolation of the NN outside its Phase 1 training envelope. What is likely occurring is as follows. TEM electron heat fluxes, when TEM is unstable, can increase with decreasing {\rlti}, as shown in the central panel of figure~\ref{fig:NN_performance_1D}. Conversely, when ITG is unstable, the ion heat flux can increase with decreasing {\rlte}. When no combined low {\rlti} and {\rlte} is available in the dataset, which typically corresponds to mode stability, then a runaway situation can occur. Decreasing {\rlte} and {\rlti} leads to increasing flux, leading to profile flattening, and further model extrapolation to yet higher fluxes. This is due to the NN extrapolating the individual heat flux trends observed in the unstable zone. In our application, due to the lattice nature of the training dataset, such extrapolations are simple to identify in applications and rectify through appropriately broadening the training set. However, in general, a method of surrogate model UQ is vital to judge the trustworthiness of model output. 

\printbibliography

\end{document}